# Widespread signals of convergent adaptation to high altitude in Asia and America

Matthieu Foll [1,2,3,*], Oscar E. Gaggiotti [4,5], Josephine T. Daub [1,2], Alexandra Vatsiou [5] and Laurent Excoffier [1,2]

[1] CMPG, Institute of Ecology and Evolution, University of Berne, Berne, 3012, Switzerland

[2] Swiss Institute of Bioinformatics, Lausanne, 1015, Switzerland

[3] Present address: School of Life Sciences, École Polytechnique Fédérale de Lausanne (EPFL), Lausanne, 1015, Switzerland

[4] School of Biology, Scottish Oceans Institute, University of St Andrews, St Andrews, Fife, KY16 8LB, UK

[5] Laboratoire d'Ecologie Alpine (LECA), UMR 5553 CNRS-Université de Grenoble, Grenoble, France

* Corresponding author: matthieu.foll@epfl.ch

# Abstract

Living at high-altitude is one of the most difficult challenges that humans had to cope with during their evolution. Whereas several genomic studies have revealed some of the genetic bases of adaptations in Tibetan, Andean and Ethiopian populations, relatively little evidence of convergent evolution to altitude in different continents has accumulated. This lack of evidence can be due to truly different evolutionary responses, but it can be also due to the low power of former studies that have mainly focused on populations from a single geographical region or performed separate analyses on multiple pairs of populations to avoid problems linked to shared histories between some populations. We introduce here a hierarchical Bayesian method to detect local adaptation that can deal with complex demographic histories. Our method can identify selection occurring at different scales, as well as convergent adaptation in different regions. We apply our approach to the analysis of a large SNP dataset from low- and high-altitude human populations from America and Asia. The simultaneous analysis of these two geographic areas allows us to identify several candidate genome regions for altitudinal selection, and we show that convergent evolution among continents has been quite common. In addition to identifying several genes and biological processes involved in high altitude adaptation, we identify two specific biological pathways that could have evolved in both continents to counter toxic effects induced by hypoxia.

# Introduction

Distinguishing between neutral and selected molecular variation has been a long-standing interest of population geneticists. This interest was fostered by the publication of Kimura's seminal paper [1] on the neutral theory of molecular evolution. Although the controversy rests mainly on the relative importance of genetic drift and selection as explanatory processes for the observed biodiversity patterns, another important question concerns the prevalent form of natural selection. Kimura [1] argued that the main selective mechanism was negative selection against deleterious mutations. However, an alternative point of view emphasizes the prevalence of positive selection, the mechanism that can lead to local adaptation and eventually to speciation [2,3].

A powerful approach to uncover positive selection is the study of mechanisms underlying convergent evolution. When different populations or evolutionary lineages are exposed to similar environments, positive selection should indeed lead to similar phenotypic features. Convergent evolution can be achieved through similar genetic changes (sometimes called "parallel evolution") at different levels: the same mutation appearing independently in different populations, the same existing mutation being recruited by selection in different populations, or the involvement of different mutations in the same genes or the same biological pathways in separate populations [4]. However, existing statistical genetic methods are not well adapted to the study of convergent evolution when data sets consists in multiple contrasts of populations living in different environments [5]. The current strategy is to carry out independent genome scans in each geographic region and to look for overlaps between loci or pathways that are identified as outliers in different regions [6]. Furthermore, studies are often split into a series of pairwise analyses that consider sets of populations inhabiting different environments. Whereas this strategy has the advantage of not requiring the modeling of complex demographic histories [7,8], it often ignores the correlation in gene frequencies between geographical regions when correcting for multiple tests [9]. As a consequence, current approaches are restricted to the comparison of lists of candidate SNPs or genomic regions obtained from multiple pairwise comparisons. This sub-optimal approach may also result in a global loss of power as compared to a single global analysis and thus to a possible underestimation of the genome-wide prevalence of convergent adaptation.

One particularly important example where this type of problems arises is in the study of local adaptation to high altitude in humans. Human populations living at high altitude need to cope with one of the most stressful environment in the world, to which they are likely to have developed specific adaptations. The harsh conditions associated with high altitude include not only low oxygen partial pressure, referred to as high-altitude hypoxia, but also other factors like low temperatures, arid climate, high solar radiation and low soil quality. While some of these stresses can be buffered by behavioral and cultural adjustments, important physiological changes have been identified in populations living at high altitude (see below). Recently, genomic advances have unveiled the first genetic bases of these physiological changes in Tibetan, Andean and Ethiopian populations [10-19]. The study of convergent or independent adaptation to altitude is of primary interest [11,19,20], but this problem has been superficially addressed so far, as most studies focused on populations from a single geographical region [10,13,14,16-19].

Several candidate genes for adaptation to altitude have nevertheless been clearly identified [21,22], the most prominent ones being involved in the hypoxia inducible factor (HIF) pathway, which plays a major role in response to hypoxia [23]. In Andeans, *VEGFA* (vascular endothelial growth factor A, MIM 192240), *PRKAA1* (protein kinase, AMP-activated, alpha 1 catalytic subunit, MIM 602739) and *NOS2A* (nitric oxide synthase 2A, MIM 163730) are the best-supported candidates, as well as *EGLN1* (egl-9 family hypoxia-inducible factor 1, MIM 606425), a down regulator of some HIF targets [12,24]. In Tibetans [10,11,13,14,16,25], the HIF pathway gene *EPAS1* (endothelial PAS domain protein 1, MIM 603349) and *EGLN1* have been repeatedly identified [22]. Recently, three similar studies focused on Ethiopian highlanders [17-19] suggested the involvement of HIF genes other than those identified in Tibetans and Andeans, with *BHLHE41* (MIM 606200), *THRB* (MIM 190160), *RORA* (MIM 600825) and *ARNT2* (MIM 606036) being the most prominent candidates.

However, there is little overlap in the list of significant genes in these three regions [18,19], with perhaps the exception of alcohol dehydrogenase genes identified in two out of the three analyses. Another exception is *EGLN1*: a comparative analysis of Tibetan and Andean populations [12] concluded that “the Tibetan and Andean patterns of genetic adaptation are largely distinct from one another“, identifying a single gene (*EGLN1*) under convergent evolution, but with both populations exhibiting a distinct dominant haplotype around this gene. This limited convergence does not contradict available physiological data, as Tibetans exhibit some phenotypic traits that are not found in Andeans [26]. For example, Tibetan populations show lower hemoglobin concentration and oxygen saturation than Andean populations at the same altitude [27]. Andeans and Tibetans also differ in their hypoxic ventilatory response, birth weight and pulmonary hypertension [28]. Finally, *EGLN1* has also been identified as a candidate gene in Kubachians, a high altitude (~2000 m a. s. l.) Daghestani population from the Caucasus [15], as well as in Indians [29].

Nevertheless, it is still possible that the small number of genes under convergent evolution is due to a lack of power of genome scan methods done on separate pairs of populations. In order to overcome these difficulties, we introduce here a Bayesian genome scan method that (i) extends the F-model [30,31] to the case of a hierarchically subdivided population consisting of several migrant pools, and (ii) explicitly includes a convergent selection model.

We apply this approach to find genes, genomic regions, and biological pathways that have responded to convergent selection in the Himalayas and in the Andes.

# Material and methods

## Hierarchical Bayesian model

One of the most widely used statistics for the comparison of allele frequencies among populations is $F_{ST}$ [32,33], and most studies cited in the introduction used it to compare low- and high altitude populations within a given geographical region (Tibet, the Andes or Ethiopia). Several methods have been proposed to detect loci under selection from $F_{ST}$, and one of the most powerful approach is based on the F-model (reviewed by Gaggiotti and Foll [34]). However, this approach assumes a simple island model where populations exchange individuals through a unique pool of migrants. This assumption is strongly violated when dealing with replicated pairs of populations across different regions, which can lead to a high rate of false positives [35].

In order to relax the rather unrealistic assumption of a unique and common pool of migrants for all sampled populations, we extended the genome scan method first introduced by Beaumont and Balding [30] and later improved by Foll and Gaggiotti [31]. More precisely, we posit that our data come from $G$ groups (migrant pools or geographic regions), each group $g$ containing $J_g$ populations. We then describe the genetic structure by a F-model that assumes that allele frequencies at locus $i$ in population $j$ from group $g$, $\boldsymbol{p_{ijg}} = \left\{ p_{ijg1}, p_{ijg2}, \ldots, p_{ijgK_i} \right\}$ (where $K_i$ is the number of distinct alleles at locus $i$), follow a Dirichlet distribution parameterized with group-specific allele frequencies $\boldsymbol{p_{ig}} = \left\{ p_{ig1}, p_{ig2}, \ldots, p_{igK_i} \right\}$ and with $F_{SC}^{ijg}$ coefficients measuring the extent of genetic differentiation of population $j$ relative to group $g$ at locus $i$. Similarly, at a higher group level, we consider an additional F-model where allele frequencies $\boldsymbol{p_{ig}}$ follow a Dirichlet distribution parameterized with meta-population allele frequencies $\boldsymbol{p_i} = \left\{ p_{i1}, p_{i2}, \ldots, p_{iK_i} \right\}$ and with $F_{CT}^{ig}$ coefficients measuring the extent of genetic differentiation of group $g$ relative to the meta-population as a whole at locus $i$. Figure S1 shows the hierarchical structure of our model in the case of three groups ($G = 3$) and four populations per group ( $J_1 = J_2 = J_3 = 4$ ) and Figure S2 shows the corresponding non-hierarchical F-model for the same number of populations. All the parameters of the hierarchical model can be estimated by further assuming that alleles in each population $j$ are sampled from a multinomial distribution [36]. These assumptions lead to an expression for the probability of observed allele counts $\boldsymbol{a_{ijg}} = \left\{ a_{ijg1}, a_{ijg2}, \ldots, a_{ijgK_i} \right\}$:

$$\Pr\left(\boldsymbol{a_{ijg}} \middle| \boldsymbol{p_{ijg}}, \boldsymbol{p_{ig}}, \boldsymbol{p_i}, \theta_{ijg}, \phi_{ig}\right) = \Pr\left(\boldsymbol{a_{ijg}} \middle| \boldsymbol{p_{ijg}}\right) \Pr\left(\boldsymbol{p_{ijg}} \middle| \boldsymbol{p_{ig}}, \theta_{ijg}\right) \Pr\left(\boldsymbol{p_{ig}} \middle| \boldsymbol{p_i}, \phi_{ig}\right)$$

where $\Pr\left(\boldsymbol{a_{ijg}} \middle| \boldsymbol{p_{ijg}}\right)$ is the multinomial likelihood, $\Pr\left(\boldsymbol{p_{ijg}} \middle| \boldsymbol{p_{ig}}, \theta_{ijg}\right)$ and $\Pr\left(\boldsymbol{p_{ig}} \middle| \boldsymbol{p_i}, \phi_{ig}\right)$ are Dirichlet prior distributions, $\theta_{ijg} = 1/F_{SC}^{ijg} - 1$, and $\phi_{ig} = 1/F_{CT}^{ig} - 1$. This expression can be simplified by integrating over $\boldsymbol{p_{ijg}}$ so as to obtain:

$$\Pr\left(\boldsymbol{a_{ijg}} \middle| \boldsymbol{p_{ig}}, \boldsymbol{p_i}, \theta_{ijg}, \phi_{ig}\right) = \Pr\left(\boldsymbol{a_{ijg}} \middle| \boldsymbol{p_{ig}}, \theta_{ijg}\right) \Pr\left(\boldsymbol{p_{ig}} \middle| \boldsymbol{p_i}, \phi_{ig}\right)$$

where $\Pr\left(\boldsymbol{a_{ijg}} \middle| \boldsymbol{p_{ig}}, \theta_{ijg}\right)$ is the multinomial-Dirichlet distribution [34]. The likelihood is obtained by multiplying across loci, regions and population

$$L\left(\boldsymbol{p_{ig}}, \boldsymbol{p_i}, \theta_{ijg}, \phi_{ig}\right) = \prod_{i=1}^{I}\prod_{g=1}^{G}\prod_{j=1}^{J_g} \Pr\left(\boldsymbol{a_{ijg}} \middle| \boldsymbol{p_{ig}}, \boldsymbol{p_i}, \theta_{ijg}, \phi_{ig}\right)$$

Using this model, we incorporate potential deviation from the genome wide F-statistics at each locus as in Beaumont and Balding [30]. The genetic differentiation within each group $g$ is:

$$\log\left(\frac{F_{SC}^{ijg}}{1-F_{SC}^{ijg}}\right) = \alpha_{ig} + \beta_{jg} \tag{1}$$

where $\alpha_{ig}$ is a locus-specific component of $F_{SC}^{ijg}$ shared by all populations in group $g$, and $\beta_{jg}$ is a population-specific component shared by all loci. Similarly, we decompose the genetic differentiation at the group level under a logistic model as:

$$\log\left(\frac{F_{CT}^{ig}}{1-F_{CT}^{ig}}\right) = A_i + B_g \tag{2}$$

where $A_i$ is a locus-specific component of $F_{CT}^{ig}$ shared by all groups in the meta-population, and $B_g$ is a group-specific component shared by all loci.

By doing this, our model also eliminates the ambiguity of having a single $\alpha_i$ parameter for more than two populations, since we now have (i) different selection parameters in each geographic region ($\alpha_{ig}$ are group specific) and (ii) separate parameters sensitive to adaptation among regions at the larger scale ($A_i$). We use the likelihood function and the logistic decomposition to derive the full Bayesian posterior:

$$\Pr\left(\boldsymbol{p_{ig}}, \boldsymbol{p_i}, A_{ig}, B_{jg}, \alpha_i, \beta_g \middle| \mathbf{A}\right) \propto L\left(\boldsymbol{p_{ig}}, \boldsymbol{p_i}, \theta_{ijg}, \phi_{ig}\right) \Pr\left(\boldsymbol{p_{ig}} \middle| \boldsymbol{p_i}, \alpha_i, \beta_g\right) \Pr\left(\boldsymbol{p_i}\right) \Pr\left(A_{ig}\right) \Pr\left(B_{jg}\right) \Pr\left(\alpha_i\right) \Pr\left(\beta_g\right)$$

where the prior for $\boldsymbol{p}_i$ is a non-informative Dirichlet distribution, the priors for $\alpha_{ig}$ and $A_i$ are Gaussian with mean 0 and variance 1, and the priors for $\beta_{jg}$ and $B_g$ are Gaussian with mean -1 and variance 1. Note that priors on $\beta_{jg}$ and $B_g$ have practically no influence on the posteriors as these parameter use the huge amount of information coming from all loci.

## Parameter estimation

We extend the Reversible Jump Markov Chain Monte Carlo (RJ-MCMC) approach proposed by Foll and Gaggiotti [31] to identify selection both within groups and among groups. For each locus and in each group separately, we consider a neutral model where $\alpha_{ig} = 0$, and a model with selection where the locus-specific effect $\alpha_{ig} \neq 0$. Similarly, we consider two models at the group level for each locus where $A_i = 0$ for the neutral model, and $A_i \neq 0$ for the model with selection. In order to tailor our approach to study convergent adaptation, we also consider the case where different groups share a unique locus-specific component $\alpha_i$ (see Figure 1 for an example of such a model with two groups of two populations). At each iteration of the MCMC algorithm, we update $A_i$ and $\alpha_{ig}$ in a randomly chosen group $g$ for all loci. As described in Foll and Gaggiotti [31], we propose to remove $\alpha_{ig}$ from the model if it is currently present, or to add it if it is not, and we do the same for $A_i$. We also add a specific Reversible Jump proposal for convergent adaptation: if all groups but one are currently in the selection model ($\alpha_{ig} \neq 0$ for all $g$ but one), we propose with a probability 0.5 to move to the convergent evolution model (where we replace all $\alpha_{ig}$ by a single selection parameter $\alpha_i$ shared by all groups), and with a probability 0.5 we perform a standard jump as described above.

## Genomic data set

In order to improve our understanding of the genetic bases of adaptation to altitude, we have applied our hierarchical Bayesian method to the dataset from Bigham et al. [12]. This data set consists of 906,600 SNPs genotyped in four populations using the Affymetrix Genome-Wide Human SNP Array 6.0 platform (see Web Resources). These four populations consist of two populations living at high altitude in the Andes (49 individuals) and in Tibet (49 individuals), as well as two lowland related populations from Central-America (39 Mesoamericans) and East Asia (90 individuals from the international HapMap project [37]). Thus, we compared four alternative models for each locus at the population level: 1) a neutral model ($\alpha_{i1} = \alpha_{i2} = 0$), 2) a model with selection acting only in Tibetans ($\alpha_{i2} = 0$), 3) a model with selection acting only in Andeans ($\alpha_{i1} = 0$), and 4) a convergent adaptation model with selection acting in both Tibetans and Andeans ($\alpha_{i1} = \alpha_{i2} = \alpha_i$). We estimate the posterior probability that a locus is under selection by summing up the posterior probabilities of the three non-neutral models (2, 3 and 4) and we control for False Discovery Rate (FDR) by calculating associated *q*-values [38-40], which are a Bayesian analogues of *p*-values taking into account multiple testing. For a given SNP, a *q*-value corresponds to the expected FDR if its posterior probability is used as a significance threshold. We do not pay any particular attention to the $A_i$ parameter here, as it can be interpreted as a potential adaptation at the continental level in Asians and Native Americans, which is not directly relevant in the context of adaptation to high altitude (but see Discussion).

We excluded SNPs with a global minor allele frequency below 5% to avoid potential biases due to uninformative polymorphisms [41]. This left us with 632,344 SNPs that were analyzed using the hierarchical F-model described above. We identified genomic regions potentially involved in high altitude adaptation by using a sliding windows approach. We considered windows of 500 kb, with a shifting increment of 25 kb at each step. The average number of SNPs per window over the whole genome was 121.4 (sd=44.6), after discarding any window containing less than 50 SNPs. We considered a window as a candidate target for selection if the 5% quantile of the *q*-values in the window was lower than 0.01, and we merged overlapping significant windows into larger regions.

## Detecting polygenic convergent adaptation

We first used SNPs identified as being under convergent adaptation to perform classical enrichment tests for pathways using Panther (see Web Resources) [42] and Gene Ontology (GO) [43] using String 9.1 (see Web Resources) [44]. More specifically, we extracted the list of 302 genes within 10 kb of all SNPs assigned to the convergent adaptation model and showing a *q*-value below 10%, to serve as input for these tests.

These two approaches have limited power to detect selection acting on polygenic traits, for which adaptive events may have arisen from standing variation rather than from new mutations [3,25]. In order to detect polygenic adaptation, we used a recent gene set enrichment approach [45], which tests if the distribution of a statistic computed across genes of a given gene set is significantly different from the rest of the genome. As opposed to the classical enrichment tests, this method does not rely on an arbitrary threshold to define the top outliers and it uses all genes that include at least one tested SNP. In short, we tested more than 1,300 gene sets listed in the Biosystems database [46] for a significant shift in their distribution of selection scores relative to the baseline genome-wide distribution. In our case, the selection score of each SNP is its associated *q*-value of convergent selection. As previously done [45], we calculated the sum of gene scores for each gene set and compared it to a null distribution of random sets (N=500,000) to infer its significance (see "Gene set enrichment analysis method" section in the Appendix). In order to avoid any redundancy between gene sets, we iteratively removed genes belonging to the most significant gene sets from the less significant gene sets before testing them again in a process called "pruning". This process leads to a list of gene sets whose significance is obtained by testing completely non-overlapping lists of genes. See the Appendix for a more detailed description of the method.

## Independent SNP Simulations

In order to evaluate the performance of our hierarchical method, we simulated data with features similar to the genomic data set analyzed here under our hierarchical F-model. Our simulated scenario thus includes two groups of two populations made of 50 diploids each, with $F_{SC} = 0.02$ for all four populations and $F_{CT} = 0.08$ for both groups. Note that these F-statistics correspond to those measured on the genomic data set we have analyzed here. In each group, a fraction of loci are under selective pressure in one of the two populations only. We simulated a total of 100,000 independent SNPs among which (i) 2,500 are under weak convergent evolution with $\alpha_i = 3$, (ii) 2,500 are under stronger convergent evolution with $\alpha_i = 5$, (iii) 2,500 are under weak selection in the first group with $\alpha_{i1} = 3$ and neutral ($\alpha_{i2} = 0$) in the second group, (iv) 2,500 are under stronger selection in the first group with $\alpha_{i1} = 5$ and neutral ($\alpha_{i2} = 0$) in the second group, and (v) 90,000 remaining SNPs that are completely neutral ($\alpha_{i1} = \alpha_{i2} = 0$). As in the real data, we conditioned the SNPs to have a global minor allele frequency above 5%. We analyzed this simulated dataset using three different approaches: (i) the hierarchical F-model introduced above, (ii) two separate pairwise analyses (one for each group) containing two populations using the original F-model implemented in BayeScan [31] (see Web Resources), (iii) a single analysis containing the four populations using the original F-model implemented in BayeScan [31] ignoring the hierarchical structure of the populations. In our hierarchal model, the best selection model for each SNP was identified as described above using a *q-value<0.01*. When analyzing data in separate pairs of populations, we considered a SNP to be under convergent adaptation when it had a *q-value<0.01* in the two regions.

## Haplotype-based simulations and statistics

Several alternative methods exist to detect natural selection. In particular, methods based on haplotype structure [47-51] have been widely applied to identify local adaptation to high altitude in humans (including the dataset from Bigham *et al.* [12] we are using here). In order to compare the performance of our approach with haplotype-based methods (see below), we have simulated haplotypic datasets with features similar to the genomic data set analysed here. We used the SimuPop package for Python [52] (see Web Resources) and considered a scenario where an ancestral population gives rise to two descendant populations, which after 600 generations (15,000 years) undergo separate splits into two populations, one at sea level and the other at high altitude. After the second split, populations evolve for 200 generations (5000 years) until the present time. This evolutionary scenario is supposed to approximate the divergence of Asian and Ameridian population followed by a subsequent divergence of highland and lowland population in Asia and in America, even though this history might have been more complex [53,54]. We assume that there is no migration between populations and we adjusted population sizes so that $F_{SC} = 0.02$ for all four populations and $F_{CT} = 0.08$ for both groups, to have F-statistics values comparable to the observed data set. More precisely, we used $N_e$=10,000 for the ancestral population, $N_e$=4,000 for the two descendant populations and $N_e$=3,500 for each of the four populations after the second split. Recombination rate was set to $10^{-8}$ (*=1cM/Mb*) and the mutation rate to $1.2\times10^{-8}$ [55]. We considered a strong selection scenario (*Ns*=100) and a moderate selection scenario (*Ns*=10), with positive selection operating only in high altitude populations right after the second split. We simulated 1,500 genomic regions each with 101 SNPs spaced every 4kb, of which (i) 1,000 were neutral, (ii) 250 were under moderate convergent evolution (*Ns*=10 in the two high-altitude populations), (iii) 250 were under strong convergent evolution (*Ns*=100 in the two high-altitude populations). For selected regions, selection operates on the SNP located at the center of the genomic region (i.e. SNP 50). We generated datasets that differed in the initial allele frequency (*IAF*) of the selected variant: (i) *IAF*=0.001, (ii) *IAF*=0.01, and (iii) *IAF*=0.1. At the end of the simulations, we sampled 50 individuals from each population and analysed the resulting dataset using different approaches. We used two commonly used statistics describing the pattern of long range homozygosity: the integrated haplotype score iHS based on the decay of haplotype homozygosity with recombination distance [48] and the cross-population extended haplotype homozygosity (XP-EHH), which contrasts the evidence of positive selection between two populations [49], and which is therefore particularly well suited to our case. Overall, we thus compared four different approaches: (i) the hierarchical F-model introduced above, (ii) two separate pairwise analyses (one for each group) containing two population using the original F-model implemented in BayeScan, (iii) two separate pairwise analyses (one for each group) containing two population using XP-EHH, and (iv) two separate analyses of the high altitude populations using iHS. We used receiver operating characteristic (ROC) curves and the area under the curve (AUC) to compare the performance of the four approaches as implemented in the R package pROC [56]. Except for the hierarchical F-model introduced above, none of these approaches can explicitly model convergent evolution, and convergent adaptation is only inferred after separate analyses when significance is reached in the two regions at the same time.

# Results

## Patterns of selection at the SNP level

Using our hierarchical Bayesian analysis, we identified 1,159 SNPs potentially under selection at the 1% FDR level (*q-value<0.01*). For each SNP, we identified the model of selection (selection only in Asia, selection only in South America, or convergent selection; see methods) with the highest posterior probability. With this procedure, 362 SNPs (31%) were found under convergent adaptation, whereas 611 SNPs (53%) were found under selection only in Asia, and 186 SNPs (16%) only under selection in South America. These results suggest that convergent adaptation is more common than previously thought [5,24,57], even at the SNP level, but consistent with results of a recent literature meta-analysis over several species [5].

In order to evaluate the additional power gained with the simultaneous analysis of the four populations, we performed separate analyses in the two continents using the non-hierarchical F-model [31]. These two pairwise comparisons identified 160 SNPs under selection in the Andes, and 940 in Tibet. The overlap in significant SNPs between these two separate analyzes and that under the hierarchical model is shown in Figure 2A. Interestingly, only 6 SNPs are found under selection in both regions when performing separate analyses in Asians and Amerindians. This very limited overlap persists even if we strongly relax the FDR in both pairwise analyzes: at the 10% FDR level only 13 SNPs are found under selection in both continents. These results are consistent with those of Bigham et al. [12], who analyzed both continents separately with a different statistical method based on $F_{ST}$ , and who found only 22 significant SNPs in common between the two geographic regions. It suggests that the use of intersecting sets of SNPs found significant in separate analyses is a sub-optimal strategy to study the genome-wide importance of convergent adaptation. Interestingly, 15% of the SNPs (162 SNPs, see Figure 2A) identified as under selection by our method are not identified by any separate analyses, suggesting a net gain in power for our method to detect genes under selection (as confirmed by our simulation studies below).

We examined in more detail the 362 SNPs identified as under convergent adaptation. The overlap of these SNPs (the yellow circle) with those identified by the two separate analyzes is shown in Figure 2B. As expected, the 6 SNPs identified under selection in both regions by the two separate analyses are part of the convergent adaptation set. However, we note that 272 of the SNPs in the convergent adaptation set (75%) are identified as being under selection in only one of the two regions by the separate analyzes. This suggests that although natural selection may be operating similarly in both regions, limited sample size may prevent its detection in one of the two continents.

## Genomic regions under selection

Using a sliding window approach, we find 25 candidate regions with length ranging from 500 kb to 2 Mb (Figure 3 and Table S1). Among these, 18 regions contain at least one significant SNP assigned to the convergent adaptation model, and 11 regions contained at least one 500 kb-window where the convergent adaptation model was the most frequently assigned selection model among significant SNPs (Figure 3). Contrastingly, Bigham et al. [12] identified 14 and 37 candidate 1 Mb regions for selection in Tibetans and Andeans, respectively, but none of these 1 Mb regions were shared between Asians and Amerindians. Moreover, only two of the regions previously found under positive selection in South America and only four in Asia overlap with our 25 significant regions.

As noted above, the only gene showing signs of convergent evolution found by Bigham et al. [12] is *EGLN1*, which has also been identified in several other studies (see Table 1 in Simonson et al. [22] for a review). *EGLN1* is also present in one of our 25 regions where three out of eight significant SNPs are assigned to the convergent adaptation model. We note that the significant SNPs in this region are not found in *EGLN1* directly but in two genes surrounding it (*TRIM67* [MIM 610584] and *DISC1* [MIM 605210]), as reported earlier [14,58]. The HIF pathway gene *EPAS1*, which is the top candidate in many studies [22], is also present in one of our 25 regions, where 28 of the 80 significant SNPs are assigned to the convergent adaptation model. Recently a particular 5-SNP *EPAS1* haplotype has been identified in Tibetans as being the result of introgression from Denisovans [54]. Unfortunately none of the five SNPs of interest identified in this study are present in our dataset, and additional sequencing will be required to check whether this haplotype is also present in Andeans.

Out of the 1,159 SNPs we identified above as being under one model of selection, 312 are located within our 25 regions (Table S1) where 120 of them are identified as under convergent adaptation (out of a total of 362 SNPs identified as under convergent adaptation in the whole data set, see Figure 2B). Almost all the 18 regions containing at least one significant SNP assigned to the convergent adaptation model also contain SNPs where the best-supported model is selection only in Asia, or selection only in America. However it is hard to distinguish if this reflects both convergent adaptation and region specific adaptation in the same genomic region, or simply different statistical power.

## Polygenic convergent adaptation

We identified three pathways significantly enriched for genes involved in convergent adaptation using Panther [42] after Bonferroni correction at the 5% level. These are the "*metabotropic glutamate receptor group I*" pathway, the "*muscarinic acetylcholine receptor 1 and 3*" signaling pathway, and the "*epidermal growth factor receptor*" (*EGFR*, MIM 131550) signaling pathway. Using the String 9.1 database [44], two GO terms were significantly enriched for these genes when controlling for a 5% FDR: "*ethanol oxidation*" (GO:0006069) and "*positive regulation of transmission of nerve impulse*" (GO:0051971). Using a recent and more powerful gene set enrichment approach [45], we first identified 25 gene sets with an associated *q*-value below 5% (Table S2). An enrichment map showing these sets and their overlap is presented in Figure 4. There are two big clusters of overlapping gene sets, one related to Fatty Acid Oxidation with *"Fatty Acid Omega Oxidation"* as the most significant set and another immune system related cluster with *"Interferon gamma signaling"* as the most significant gene set. After pruning, only these two above-mentioned gene sets are left with a *q*-value below 5%. It is worth noting that the "*Fatty Acid Omega Oxidation"* pathway, which is the most significant gene set ($q$-value$<10^{-6}$), contains many top scoring genes for convergent selection, including several alcohol and aldehyde dehydrogenases, as listed in Table S3. Interestingly, the GO term "*ethanol oxidation*" is no longer significant after excluding the genes involved in the *"Fatty Acid Omega Oxidation"* pathway. Out of the 362 SNPs identified under convergent adaptation above, only four are located in genes (±50kb) belonging to the *Fatty Acid Omega Oxidation* pathway (rs3805322, rs2051428, rs4767944,

rs4346023), and only seven SNPs are found in genes belonging to the *Interferon gamma signaling* pathway (rs12531711, rs7105122, rs4237544, rs10742805, rs17198158, rs4147730, rs3115628). This apparent lack of significant SNPs in candidate pathways is expected, as our gene set enrichment approach does not rely on an arbitrary threshold to define the top outliers and is thus more suited to detect lower levels of selection acting synergistically on polygenic traits.

## Power of the hierarchical F-model

Our simulations show a net increase in power to detect selection using the global hierarchical approach as compared to using two separate pairwise analyses (Table 1 and Figure 5 and 6). For the 2,500 SNPs simulated under the weak convergent selection model ( $\alpha_i = 3$ ), the hierarchical model detects 6.5 times more SNPs than the two separate analyses (306 vs. 47). This difference can be explained by the smaller amount of information used when doing separate analyses instead of a single one. The power greatly increases when selection is stronger, and among the 2,500 SNPs simulated with $\alpha_i = 5$, 1,515 are correctly classified using our hierarchical model, as compared to only 643 using separate analyses. Similarly to what we found with the real altitude data, the two separate analyses often wrongly classify the convergent SNPs correctly identified as such by our hierarchical method as being under selection in only one of the two groups, but sometimes also as completely neutral (64 such SNPs when $\alpha_i = 3$ and 76 when $\alpha_i = 5$, see Figure 5B and D). We note that the hierarchical model is also more powerful at detecting selected loci regardless of whether or not the SNPs are correctly assigned to the convergent evolution set. Indeed, our method identifies 2,626 SNPs as being under any model of selection (*i. e.* convergent evolution or in only one of the two regions) among the 5,000 simulated under convergent selection, whereas the separate analysis detects only 2,475 SNPs. When selection is present only in one of the two groups ($\alpha_{i1} = 3$ or 5 and $\alpha_{i2} = 0$), the power of the hierarchical model is comparable with the separate analysis in the corresponding group, implying that there is no penalty in using the hierarchical model even in presence of group specific selection. A few of the group-specific selected SNPs are wrongly classified in the convergent adaptation model with a false positive rate of 1.7% (84 SNPs out of 5,000). Overall, the false discovery rate is well calibrated using our *q*-value threshold of 0.01 in both cases, with 29 false positives out of 4,141 significant SNPs (FDR=0.70%) for our hierarchical model, and 30 false positives out of 3,984 significant SNPs (FDR=0.75%) for the two separate analyses. Finally, when the four populations are analyzed together without accounting for the hierarchical structure, a large number of false positives appears (Table 1 and Figure 6C) in keeping with previous studies [35]. Under this island model, 1,139 neutral SNPs are indeed identified as being under selection among the 90,000 simulated neutral SNPs (vs. 29 and 30 using the hierarchical method or two separate analyses, respectively). The non-hierarchical approach does not allow one to distinguish different models of selection, but among the 10,000 SNPs simulated under different types of selection, only 2,598 are significant. This shows that the non-hierarchical analysis leads to both a reduced power, and a very large false discovery rate (FDR=30.4%) in presence of a hierarchical genetic structure.

Our haplotype-based simulations also show that our hierarchical model has generally a much higher performance than iHS and XP-EHH to detect convergent adaptation (Table 2). iHS has very low power to detect selection in all scenarios tested here, while XP-EHH performs well (*AUC*=0.75) when *IAF*=0.1 and selection is moderate (*Ns*=10), and very well (*AUC*=0.94) when *IAF*=0.001 and selection is strong (*Ns*=100). ROC curves for these two cases are presented in Figure 7, and for the four other cases in Figure S3. In both of these cases however, the hierarchical model has a higher performance (*AUC*=0.92 and *AUC=0*.999, respectively), and it also shows a high performance (*AUC*=0.92 and *AUC*=0.94) in the two other scenarios with strong selection (*Ns*=100) where XP-EHH has almost no power to detect convergent adaptation (*AUC*=0.57 and *AUC*=0.50 respectively, see Table 2). Interestingly, for the case where *IAF*=0.1 and *Ns*=10*,* XP-EHH has a slightly higher performance to detect selection in each region individually than the F-model as implemented in BayeScan (*AUC*=0.75 vs. *AUC*=0.71, respectively), but our hierarchical model outperforms these two approaches drastically (*AUC*=0.92, Figure 7A). Note that XP-EHH performs somewhat better than the other methods in one scenario (*Ns*=10, *IAF*=0.01), but its performance (*UAC*=0.60) is not particularly good and all methods seem to have problems to detect convergent adaptation in this case. Overall our analyses confirm that the use of separate analyses results in reduced power to detect convergent adaptation, which explains the difference between results obtained using our and previous methods when detecting high altitude adaptation in humans. The ROC analysis also shows that using a less stringent cutoff in separate analyses is far from being as powerful as our hierarchical model.

# Discussion

## Convergent adaptation to high altitude in Asia and America is not rare

Our hierarchical F-model reveals that convergent adaptation to high altitude is more frequent than previously described in Tibetans and Andeans. Indeed, 31% (362/1159) of all SNPs found to be potentially under selection at a FDR of 1% can be considered as under convergent adaptation in Asia and America. This is in sharp contrast with a previous analysis of the same data set where only a single gene was found to be responding to altitudinal selection in both Asians and Amerindians [12]. Our model confirms the selection of *EGLN1* in both Tibetans and Andeans. We also show that some genes already known to be involved in adaptation to high altitude in Tibetans, like *EPAS1*, may also have the same function in Andeans. Finally, we identified genomic regions, pathways, and GO terms potentially linked to convergent adaptation to high altitude in Tibetans and Andeans that have not been previously reported. Our approach seems thus more powerful than previous pairwise analyses, which is confirmed by our simulation studies. It suggests that datasets analyzed by previous studies that tried to uncover convergent adaptation by confronting lists of significant SNPs in separate pairwise analyses [59-63] would benefit from being reanalyzed with our method. We note that more complex demography could lead to a false positive rate higher than the nominal value. Based on the simple scenario of the divergence of four populations we have simulated, we found that our method is robust to the assumed demographic model, but this may not be always the case, and significant SNPs have to be considered only as candidates for further investigations.

## Polygenic and convergent adaptation in the omega oxidation pathway

Our top significant GO term is linked to alcohol metabolism, in keeping with a recent study of a high altitude population in Ethiopia [18,19]. Indeed, one of the 25 regions identified in the present study includes several alcohol dehydrogenase (*ADH*) genes (*ADH1A* [MIM 103700], *ADH1B* [MIM 103720], *ADH1C* [MIM 103730], *ADH4* [MIM 103740], *ADH5* [MIM 103710], *ADH6* [MIM 103735], *ADH7* [MIM 600086]) located in a 370 kb segment of chromosome 4 (Figure 3), and another significant segment of 2 Mb portion of chromosome 12 includes *ALDH2* (acetaldehyde dehydrogenase, MIM 100650). Some evidence of positive selection in *ADH1B* and *ALDH2* had been reported in East-Asian populations, but without any clear selective forces identified [64].

Interestingly, our gene set enrichment analysis suggests a potential evolutionary adaptation of this group of genes, since they all belong to our most significant pathway, namely "*Fatty Acid Omega Oxidation*" (Table S3). Omega oxidation is an alternative to the beta-oxidation pathway involved in fatty acid degradation and energy production in the mitochondrion. Degradation of fatty acids into sugar by omega oxidation is usually a minor metabolic pathway, which becomes more important when beta-oxidation is defective [18,19,65], or in case of hypoxia [66]. It is however unclear if omega oxidation is a more efficient alternative to beta oxidation at high altitude, or if it would rather contribute to the degradation of fatty acids accumulating when beta oxidation is defective. The detoxifying role of this pathway is supported by the fact that it is usually mainly active in the liver and in the kidney [65]. The fact that Ethiopians also show signals of adaptations in *ADH* and *ALDH* genes [19] suggests that convergent adaptation in the omega oxidation pathway could have occurred on three different continents in humans.

## Response to hypoxia-induced neuronal damage

Hypoxia leads to neuronal damage through over-stimulation of glutamate receptors [67].Two out of our three significant pathways found with Panther (“*metabotropic glutamate receptor group I*” and “*muscarinic acetylcholine receptor 1 and 3*”) for convergent adaptation are involved with neurotransmitter receptors. The metabotropic glutamate receptor group I increases N-methyl-D-aspartate (NMDA) receptor activity, and this excitotoxicity is a major mechanism of neuronal damage and apoptosis [68]. Consistently, the only significant GO term after excluding the genes involved in omega oxidation is also related to neurotransmission (“*positive regulation of transmission of nerve impulse*”) and contains two significant glutamate receptors genes (*GRIK2* [MIM 138244] and *GRIN2B* [138252]) as well as *IL6* (MIM 147620).

One of our top candidate regions for convergent adaptation includes 19 significant SNPs assigned to the convergent adaptation model, which are spread in a 100 kb region on chromosome 7 around *IL6* (Figure 3), the gene encoding interleukin-6 (IL-6), an important cytokine. Interestingly it has been shown that IL-6 plasma levels increases significantly when sea-level resident individuals are exposed to high altitude (4300 m) [69], and IL-6 has been shown to have a neuroprotective effect against glutamate- or NMDA-induced excitotoxicity [70]. Consistently the “*metabotropic glutamate receptor group III*” pathway seems to have responded to selection in Ethiopian highlanders [17]. Together, these results suggest a genetic basis for an adaptive response to neuronal excitotoxicity induced by high altitude hypoxia in humans.

## Versatility of the hierarchical Bayesian model to uncover selection

Our statistical model is very flexible and can cope with a variety of sampling strategies to identify adaptation. For example, Pagani et al. [15] used a different sampling scheme to uncover high altitude adaptation genes in North-Caucasian highlanders. They sampled Daghestani from three closely related populations (Avars, Kubachians, and Laks) living at high altitude that they compared with two lowland European populations. Here again, our strategy would allow the incorporation of these five populations into a single analysis. A first group would correspond to the Daghestan region, containing the three populations and a second group containing the two lowland populations. However, in that case, it is the decomposition of $F_{CT}$ in equation 2 that would allow the identification of loci overly differentiated between Daghestani ("group 1") and European ("group 2") populations.

Our approach could also be very useful in the context of Genome Wide Association Studies (GWAS) meta-analysis. For example, Scherag et al. [71] combined two GWAS on French and German samples to identify SNPs associated with extreme obesity in children. These two data sets could be combined and a single analysis could be performed under our hierarchical framework, explicitly taking into account the population structure. Our two "groups" in Figure 1 would correspond respectively to French and German individuals. In each group the two "populations" would correspond respectively to cases (obese children) and controls (children with normal weight). Like in the present study, the decomposition of $F_{SC}$ and the use of a convergent evolution model would allow the identification of loci associated with obesity in both populations. Additionally, a potential hidden genetic structure between cases and controls and any shared ancestry between French and Germans would be dealt with by the $\beta_{jg}$ and $B_g$ coefficients in equations 1 and 2, respectively.

We have introduced here a flexible hierarchical Bayesian model that can deal with complex population structure, and which allows the simultaneous analysis of populations living in different environments in several distinct geographic regions. Our model can be used to specifically test for convergent adaptation, and this approach is shown to be more powerful than previous methods that analyze pairs of populations separately. The application of our method to the detection of loci and pathways under selection reveals that many genes are under convergent selection in the American and Tibetan highlanders. Interestingly, we find that two specific pathways could have evolved to counter the toxic effects of hypoxia, which adds to previous evidence (e.g. *EPAS* and *EGLN1* [22]) suggesting that human populations living at high altitude might have mainly evolved ways to limit the negative effects of normal physiological responses to hypoxia, and might not have had enough time yet to develop more elaborate adaptations to this harsh environment.

## Supplemental Data Description

Supplemental Data includes three figures and three tables.

## Acknowledgements

We thank Prof. Abigail Bigham for making the genetic data analyzed here available. This work has been made possible by Swiss NSF grants No. 3100A0-126074, 31003A-143393, and CRSII3_141940 to LE. OEG was supported by French ANR grant No 09-GENM-017-001 and by the Marine Alliance for Science and Technology for Scotland (MASTS). The program BayeScan3 used to analyze the data is available from MF upon request.

## Appendix

### Gene set enrichment analysis method

To find signals of selection at the pathway level we applied a gene set enrichment approach as described by Daub et al. [45]. This method tests whether the genes in a gene set show a shift in the distribution of a selection score. In our case we take as selection score $s_{conv}$= 1-$q_{conv}$, where $q_{conv}$ is the q-value of a SNP computed from the probability of convergent selection. For the enrichment test we need one $s_{conv}$ value per gene, we therefore transformed the SNP based scores to gene based scores. We first downloaded 19,683 protein coding human genes, located on the autosomes and on the X chromosome, from the NCBI Entrez Gene website [72] (see Web Resources). Next we converted the SNPs to hg19 coordinates. 670 SNPs could not be mapped, resulting in 631,674 remaining SNPs. These SNPs were assigned to genes: if a SNP was located within a gene transcript, it was assigned to that gene; otherwise it was assigned to the closest gene within 50kb distance. For each gene, we selected the highest $s_{conv}$ value of all SNPs assigned to this gene. After removing 2,411 genes with no SNPs assigned, a list of 17,272 genes remained.

We downloaded 2,402 gene sets from the NCBI Biosystems database [46] (see Web Resources). After discarding genes that were not part of the aforementioned gene list, removing gene sets with less than 10 genes and pooling (nearly) identical gene sets, 1,339 sets remained that served as input in our enrichment tests.

We computed the SUMSTAT [73] score for each set, which is the sum of the $s_{conv}$ values of all genes in a gene set. Gene sets with a high SUMSTAT score are likely candidates for convergent selection. To assess the significance of each tested gene set, we compared its SUMSTAT score with a null distribution of SUMSTAT scores from random gene sets (N=500,000) of the same size. We could not approximate the null distribution with a normal distribution as applied in Daub et al. [45], as random gene sets of small to moderate size produced a skewed SUMSTAT distribution. Taking the highest $s_{conv}$ score among SNPs near a gene can induce a bias, since genes with many SNPs are more likely to have an extreme value assigned. To correct for this possible bias we placed each gene in a bin containing all genes with approximately the same number of SNPs and constructed the random gene sets in the null distribution in such a way that they were composed of the same number of genes from each bin as the gene set being tested. To remove overlap among the candidate gene sets, we applied a pruning method where we assign iteratively overlapping genes to the highest scoring gene set. As these tests are not independent anymore, we empirically estimated the q-value of these pruned sets. All sets that scored a q-value <5% (before and after pruning) were reported.

## Web Resources

The URLs for data presented herein are as follows:

Affymetrix Genome-Wide Human SNP Array 6.0 description, http://www.affymetrix.com/estore/catalog/131533/AFFY/Genome-Wide+Human+SNP+Array+6.0#1_1

BayeScan version 2.1, http://cmpg.unibe.ch/software/BayeScan/

NCBI Biosystems, http://www.ncbi.nlm.nih.gov/biosystems

NCBI Entrez Gene, http://www.ncbi.nlm.nih.gov/gene

Online Mendelian Inheritance in Man (OMIM), http://www.omim.org/

PANTHER, http://www.pantherdb.org

simuPOP, http://simupop.sourceforge.net

STRING version 9.1, http://string-db.org

# Figure Titles and Legends

**Figure 1.** Hierarchical F-model for the high altitude data analyzed.

Directed acyclic graph describing the Bayesian formulation of the hierarchical F-model at a given locus $i$. Square nodes represent data and circles represent model parameters to be estimated. Dashed circles represent population allele frequencies, which are analytically integrated using a Dirichlet-multinomial distribution (see method description). Lines between the nodes represent direct stochastic relationships within the model. With the exception of Figure 4, we use the same color codes in all Figures, with blue for Asia, red for America, and yellow for convergent adaptation.

**Figure 2.** Overlap of candidate SNPs under selection in Asia and in America.

Venn diagrams showing the overlap of SNPs potentially under selection in Asia and in America at a 1% FDR.  A: Overlap between all SNPs found under any type of selection using our hierarchical model introduced here (green) with those found in separate analyses performed in Asia (blue) and in America (red). B: Overlap between SNPs found under convergent selection using our hierarchical model (yellow) with those found in separate analyses performed in Asia (blue) and in America (red).

**Figure 3.** Manhattan plot of *q*-values for loci potentially under altitudinal selection in Asian and Amerindian populations.

Each dot represents the 5% quantile of the SNPs *q*-values in a 500 kb window. Windows are shifted by increment of 25 kb and considered as a candidate target for selection if the 5% quantile is lower than 0.01 (horizontal dashed line). Overlapping significant windows are merged into 25 larger regions (indicated by grey vertical bars, see Table S2). Significant windows are colored in yellow when they contain at least one significant SNP for convergent adaptation. Otherwise they are colored according to the most represented model of selection identified among the SNPs they contain: blue for selection only in Asia and red for selection only in America. We also report the names of genes discussed in the text.

**Figure 4.** Gene sets enriched for signals of convergent adaptation.

The 25 nodes represent gene sets with *q-value<0.05*. The size of a node is proportional to the number of genes in a gene set. The node color scale represents gene set p-values. Edges represent mutual overlap: nodes are connected if one of the sets has at least 33% of its genes in common with the other gene set. The widths of the edges scale with the similarity between nodes.

**Figure 5.** Overlap of significant SNPs for simulated convergent evolution.

Venn diagrams showing the overlap of SNPs simulated under a convergent evolution model and identified under selection at a 1% FDR. A and C: Overlap between SNPs found under any type of selection using our hierarchical model introduced here (green) with those found in separate analyses performed in group 1 (blue) and in group 2 (red). B and D: Overlap between SNPs found under convergent using our hierarchical model (yellow) with those found in separate analyses performed in group 1 (blue) and in group 2 (red). In A and B, 2,500 SNPs are simulated under weak convergent selection ($\alpha_i = 3$), while in C and D 2,500 SNPs are simulated under stronger convergent selection ($\alpha_i = 5$).

**Figure 6.** Power to detect loci under selection as a function of their effect on population differentiation.

For simulated SNPs, we plot the best selection model inferred (A) under our hierarchical F-model, (B) using two separate analyses of pairs of populations, and (C) under a non-hierarchical F model performed on four populations, thus ignoring the underlying hierarchical population structure. The colors indicate the inferred model: convergent evolution (yellow), selection only in the first group (blue), selection only in the second group (red), and no selection (black). Note that we use purple in the C panel, as this approach does not allow one to distinguish between different models of selection. For better visualization, we only plot 10,000 neutral loci among the 90,000 simulated, but the missing data show a very similar pattern.

**Figure 7**. Haplotype-based simulation ROC curves.

ROC curves summarizing the relative performance of our hierarchical model, BayeScan, and XP-EHH to detect convergent adaptation for simulated scenarios when (A) *IAF*=0.1 and *Ns*=10 and (B) *IAF*=0.001 and *Ns*=100 (see also Table 2 for overall scores).

## Supplemental Data

**Figure S1**. Hierarchical F-model.

Directed acyclic graph describing the Bayesian formulation of the hierarchical F-model with 12 populations clustered in three groups at a given locus $i$. Square nodes represent data and circles represent model parameters to be estimated. Dashed circles represent population allele frequencies, which are analytically integrated using a Dirichlet-multinomial distribution (see method description). Lines between the nodes represent direct stochastic relationships within the model.

**Figure S2**. Original F-model.

Directed acyclic graph describing the Bayesian formulation of the original F-model with 12 populations at a given locus $i$. Square nodes represent data and circles represent model parameters to be estimated. Dashed circles represent population allele frequencies, which are analytically integrated using a Dirichlet-multinomial distribution (see method description). Lines between the nodes represent direct stochastic relationships within the model.

**Figure S3**. Haplotype-based simulation ROC curves.

ROC curves summarizing the relative performance of our hierarchical model, BayeScan, and XP-EHH to detect convergent adaptation for simulated scenarios (see also Table 2 for overall scores).

**Table S1**. Significant regions under altitudinal selection in Asian and Amerindian populations identified using the sliding windows approach.

The 25 genomic regions identified correspond to the vertical grey bars in Figure 3. We report the closest genes within 250kb from a significant SNP (q-value<0.01) in each region. The corresponding SNPs for each gene are also reported. We highlight in bold the genes and regions discussed in the text and in Figure 3.

**Table S2**. Gene sets enriched for signals of convergent adaptation before pruning (*q-value<0.05*).

We highlight in bold the only two gene sets that remain significant after the pruning procedure, which consists in removing overlapping genes from less significant genet sets and retesting in an iterative manner.

**Table S3**. Results of the gene set enrichment approach for the "Fatty Acid Omega Oxidation" cluster.

We report the list of genes member of all gene sets in the "Fatty Acid Omega Oxidation" cluster (see Figure 4). The only remaining significant gene set after pruning ("Fatty Acid Omega Oxidation") is highlighted in yellow. For each gene we report the retained SNP (see Appendix), the distance to the gene, and the corresponding statistic used (1-q-value).

# Tables

**Table 1**: Result of the $F_{ST}$-based simulated data analyses. We report for different methods and selection strengths the number of SNPs found to be neutral or under selection at a FDR of 1%.

| SNP category | Selection parameters | Number of SNPs | Hierarchical model | | | | Two separate BayeScan analyses | | | | Single BayeScan analysis | |
|---|---|---|---|---|---|---|---|---|---|---|---|---|
| | | | Neutral | Selection | | | Neutral | Selection | | | Neutral | Selection |
| | | | | Convergent | Group 1 | Group 2 | | Convergent | Group 1 | Group 2 | | |
| Convergent | $\alpha_i = 3$ | 2500 | 1824 | 306 | 148 | 222 | 1891 | 47 | 235 | 327 | 2127 | 373 |
| | $\alpha_i = 5$ | 2500 | 550 | 1515 | 188 | 247 | 634 | 643 | 554 | 669 | 1048 | 1452 |
| Group 1 | $\alpha_{i1} = 3$ $\alpha_{i2} = 0$ | 2500 | 2206 | 19 | 275 | 0 | 2214 | 0 | 285 | 1 | 2367 | 133 |
| | $\alpha_{i1} = 5$ $\alpha_{i2} = 0$ | 2500 | 1308 | 65 | 1127 | 0 | 1307 | 0 | 1192 | 1 | 1860 | 640 |
| Neutral | $\alpha_{i1} = \alpha_{i2} = 0$ | 90,000 | 89971 | 0 | 4 | 25 | 89,970 | 0 | 4 | 26 | 88,861 | 1139 |
| All | | 100,000 | 95859 | 1905 | 1742 | 494 | 96,016 | 690 | 2270 | 1024 | 96,263 | 3737 |

**Table 2**: Performance (AUC) of different methods to detect convergent adaptation in the case of haplotype-based simulated data sets. For the case *Ns*=100 and *IAF*=0.001 iHS could not be computed.

| | | AUC[a] | | | |
|---|---|---|---|---|---|
| **Selection strength** | **Initial allele frequency (IAF)** | **iHS** | **XP-EHH** | **BayeScan** | **Hierarchical model** |
| *Ns*=10 (moderate) | 0.001 | 0.52 | 0.54 | 0.57 | 0.59 |
| | 0.01 | 0.52 | 0.60 | 0.55 | 0.54 |
| | 0.1 | 0.54 | 0.75 | 0.71 | 0.92 |
| *Ns*=100 (strong) | 0.001 | - | 0.94 | 0.996 | 0.999 |
| | 0.01 | 0.58 | 0.57 | 0.88 | 0.94 |
| | 0.1 | 0.51 | 0.50 | 0.90 | 0.92 |

[a]AUR: area under the ROC curve (see Figure 7)

**Figure 1.**

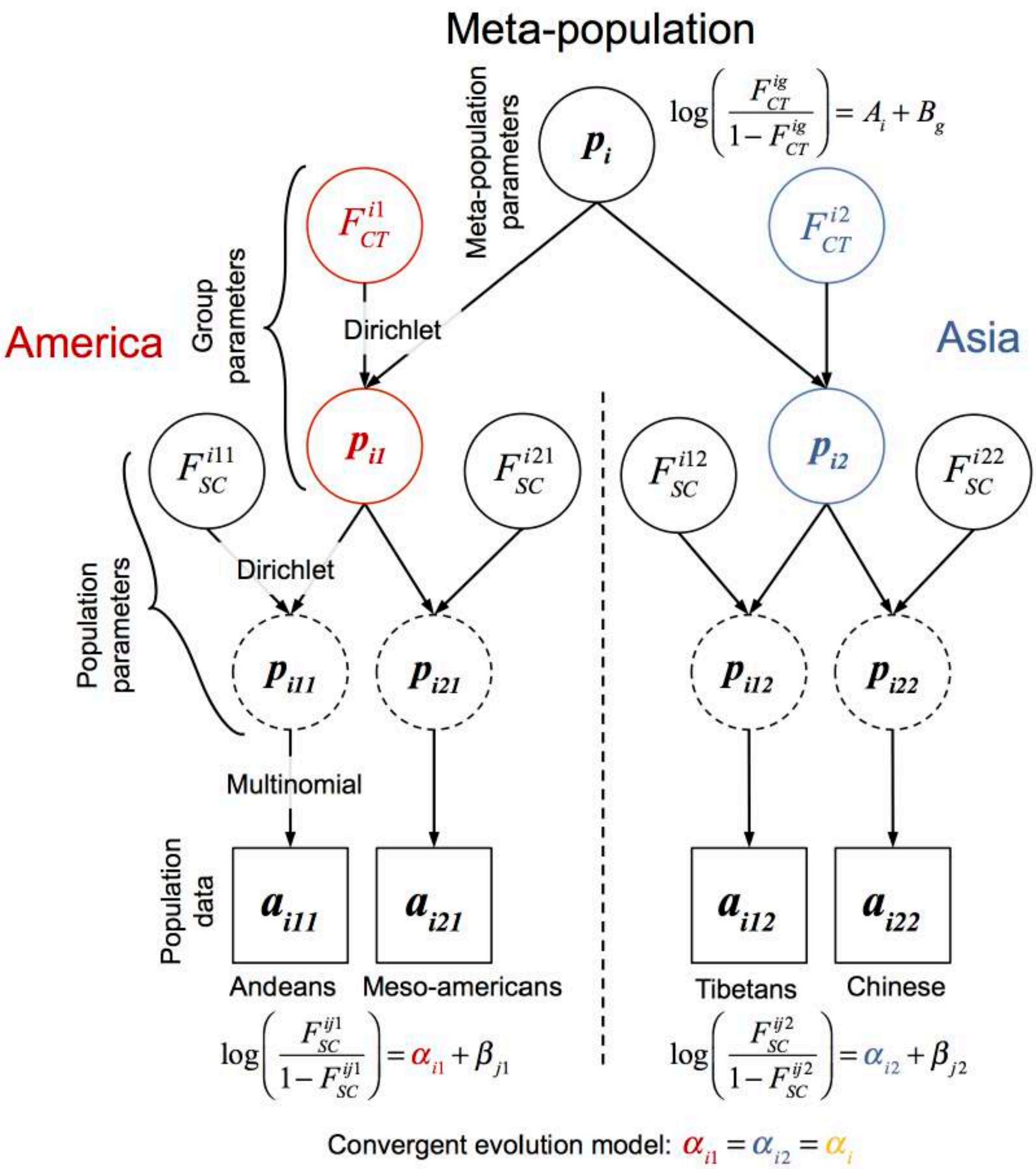

Meta-population
Meta-population parameters
$p_i$
$\log\left(\frac{F_{CT}^{ig}}{1-F_{CT}^{ig}}\right) = A_i + B_g$
America
Group parameters
$F_{CT}^{i1}$
Dirichlet
$F_{CT}^{i2}$
Asia
$p_{i1}$
$p_{i2}$
$F_{SC}^{i11}$
$F_{SC}^{i21}$
$F_{SC}^{i12}$
$F_{SC}^{i22}$
Population parameters
Dirichlet
$p_{i11}$
$p_{i21}$
$p_{i12}$
$p_{i22}$
Multinomial
Population data
$a_{i11}$
$a_{i21}$
$a_{i12}$
$a_{i22}$
Andeans
Meso-americans
Tibetans
Chinese
$\log\left(\frac{F_{SC}^{ij1}}{1-F_{SC}^{ij1}}\right) = \alpha_{i1} + \beta_{j1}$
$\log\left(\frac{F_{SC}^{ij2}}{1-F_{SC}^{ij2}}\right) = \alpha_{i2} + \beta_{j2}$
Convergent evolution model: $\alpha_{i1} = \alpha_{i2} = \alpha_i$

**Figure 2.**

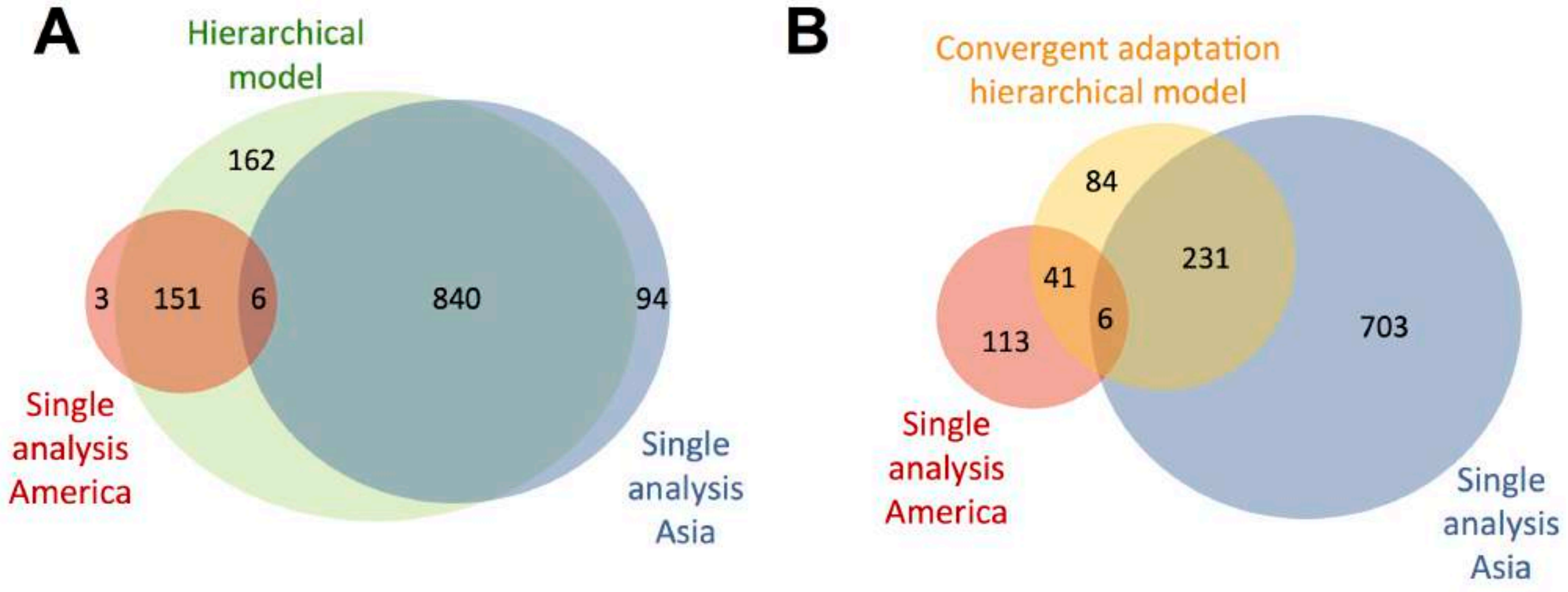

**Figure 3.**

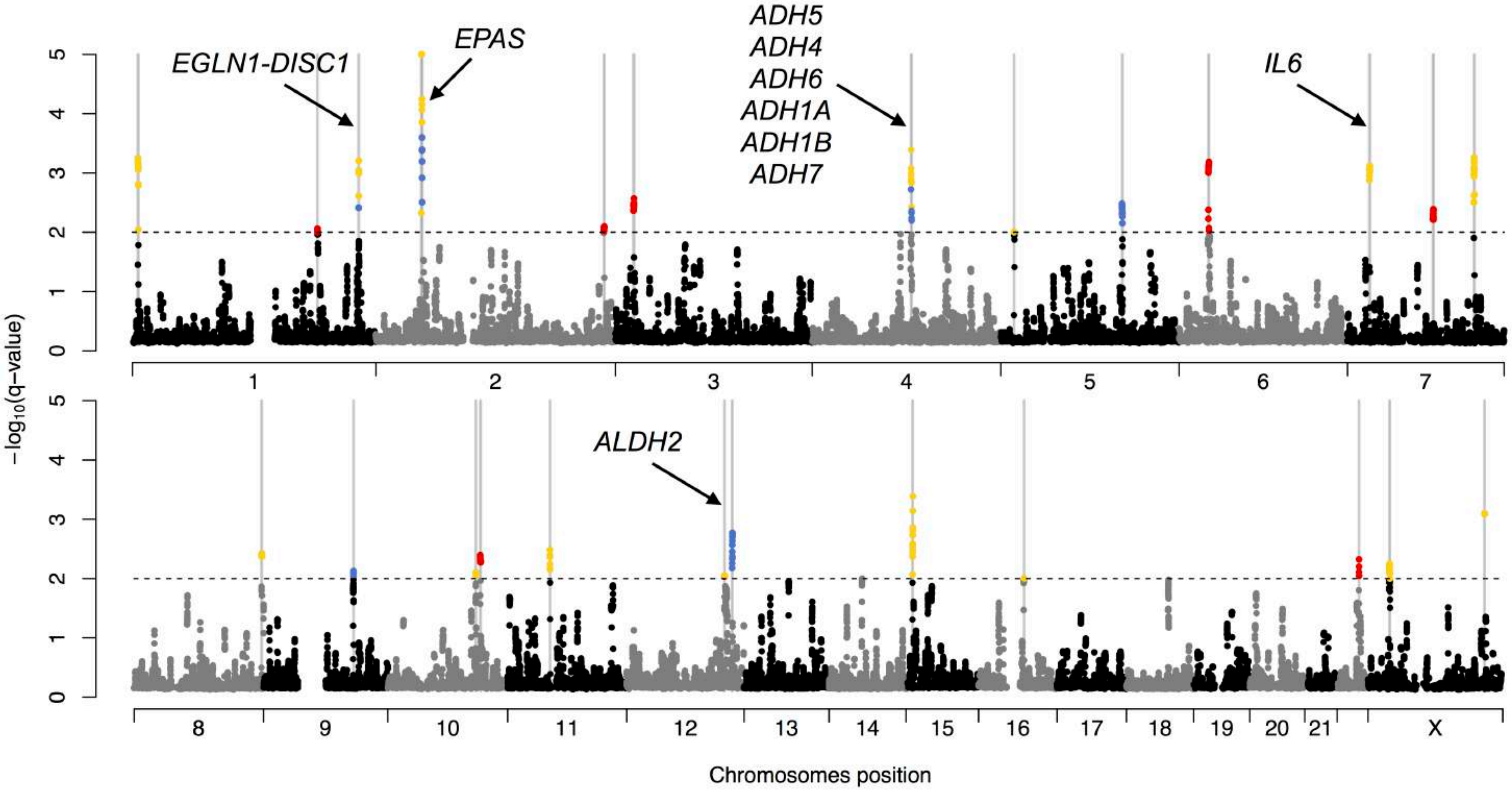

## Figure 4.

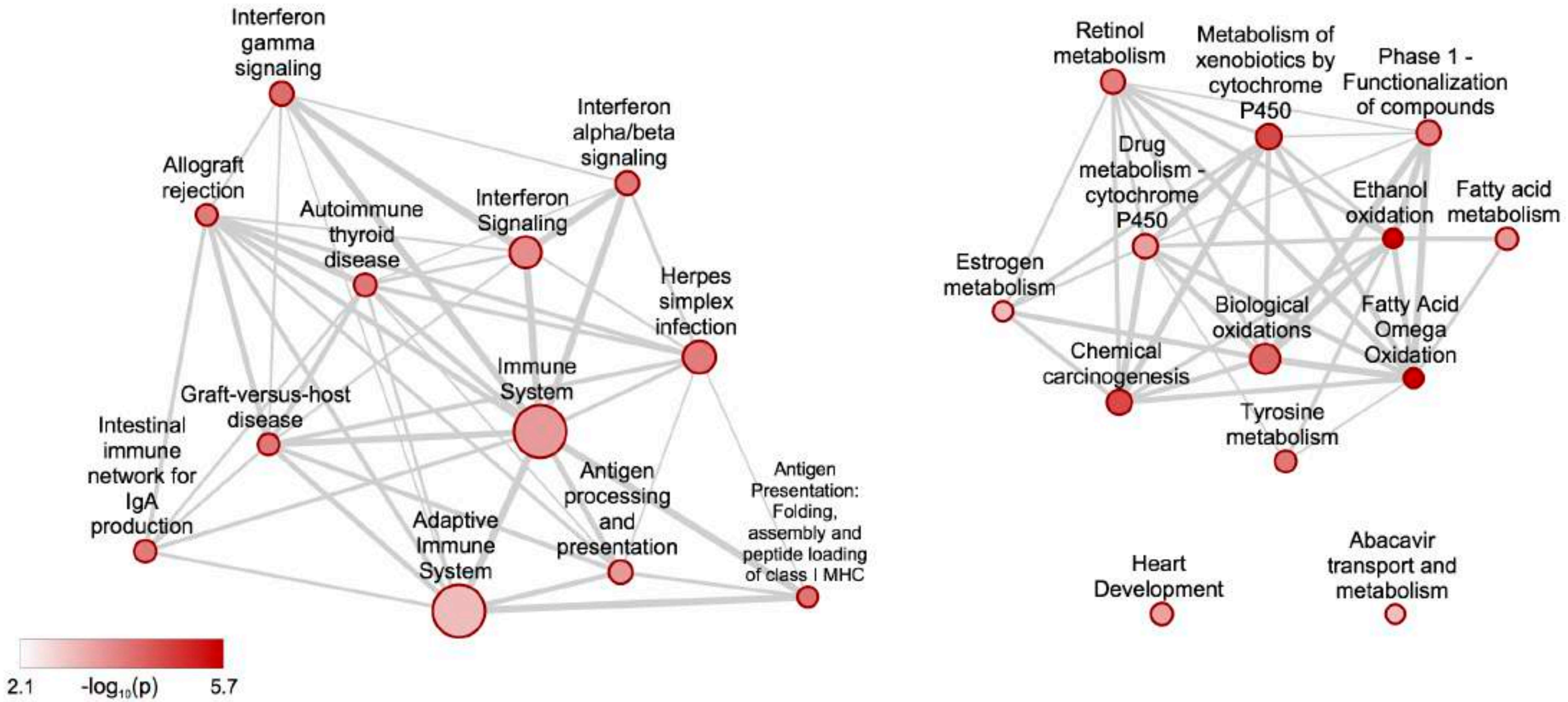
Interferon gamma signaling
Interferon alpha/beta signaling
Allograft rejection
Autoimmune thyroid disease
Interferon Signaling
Herpes simplex infection
Immune System
Graft-versus-host disease
Intestinal immune network for IgA production
Antigen processing and presentation
Antigen Presentation: Folding, assembly and peptide loading of class I MHC
Adaptive Immune System
2.1
-log10(p)
5.7
Retinol metabolism
Metabolism of xenobiotics by cytochrome P450
Phase 1 - Functionalization of compounds
Drug metabolism - cytochrome P450
Ethanol oxidation
Fatty acid metabolism
Estrogen metabolism
Biological oxidations
Fatty Acid Omega Oxidation
Chemical carcinogenesis
Tyrosine metabolism
Heart Development
Abacavir transport and metabolism

**Figure 5.**

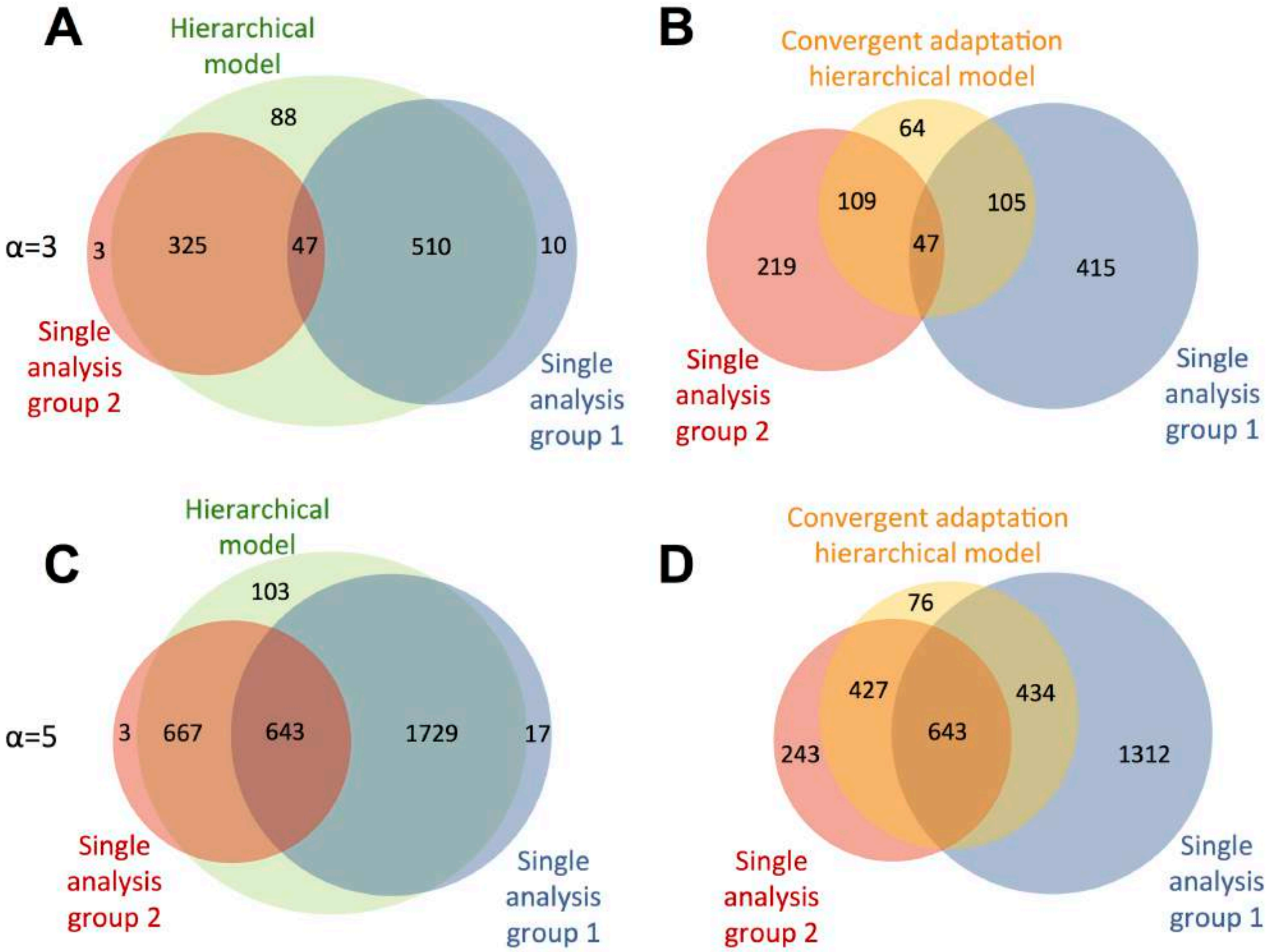
A
Hierarchical model
88
α=3
3
325
47
510
10
Single analysis group 2
Single analysis group 1
B
Convergent adaptation hierarchical model
64
109
105
47
219
415
Single analysis group 2
Single analysis group 1
C
Hierarchical model
103
α=5
3
667
643
1729
17
Single analysis group 2
Single analysis group 1
D
Convergent adaptation hierarchical model
76
427
434
643
243
1312
Single analysis group 2
Single analysis group 1

**Figure 6.**

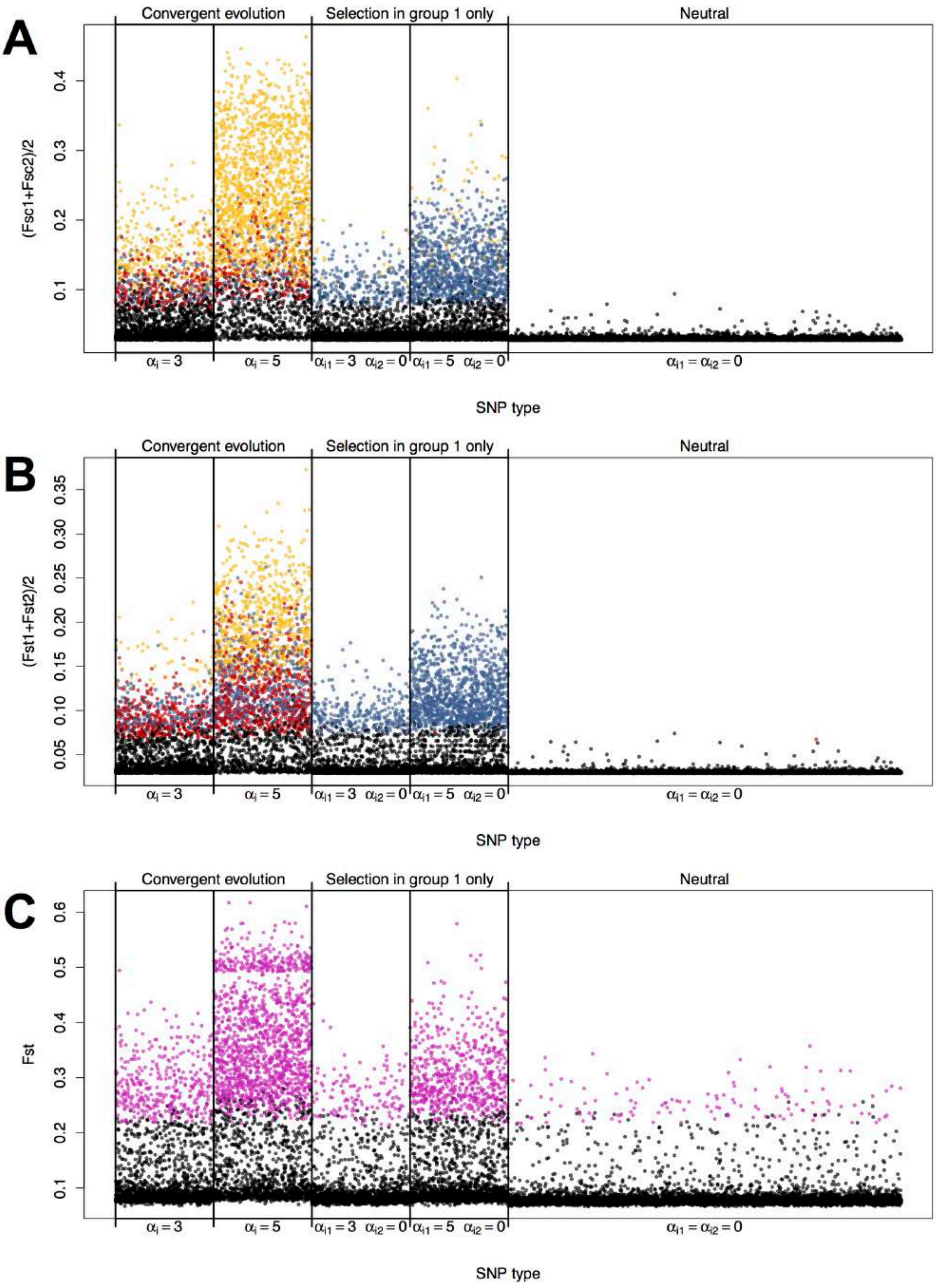
A
B
C
Convergent evolution
Selection in group 1 only
Neutral
(Fsc1+Fsc2)/2
(Fst1+Fst2)/2
Fst
αi = 3
αi = 5
αi1 = 3 αi2 = 0
αi1 = 5 αi2 = 0
αi1 = αi2 = 0
SNP type

## Figure 7.

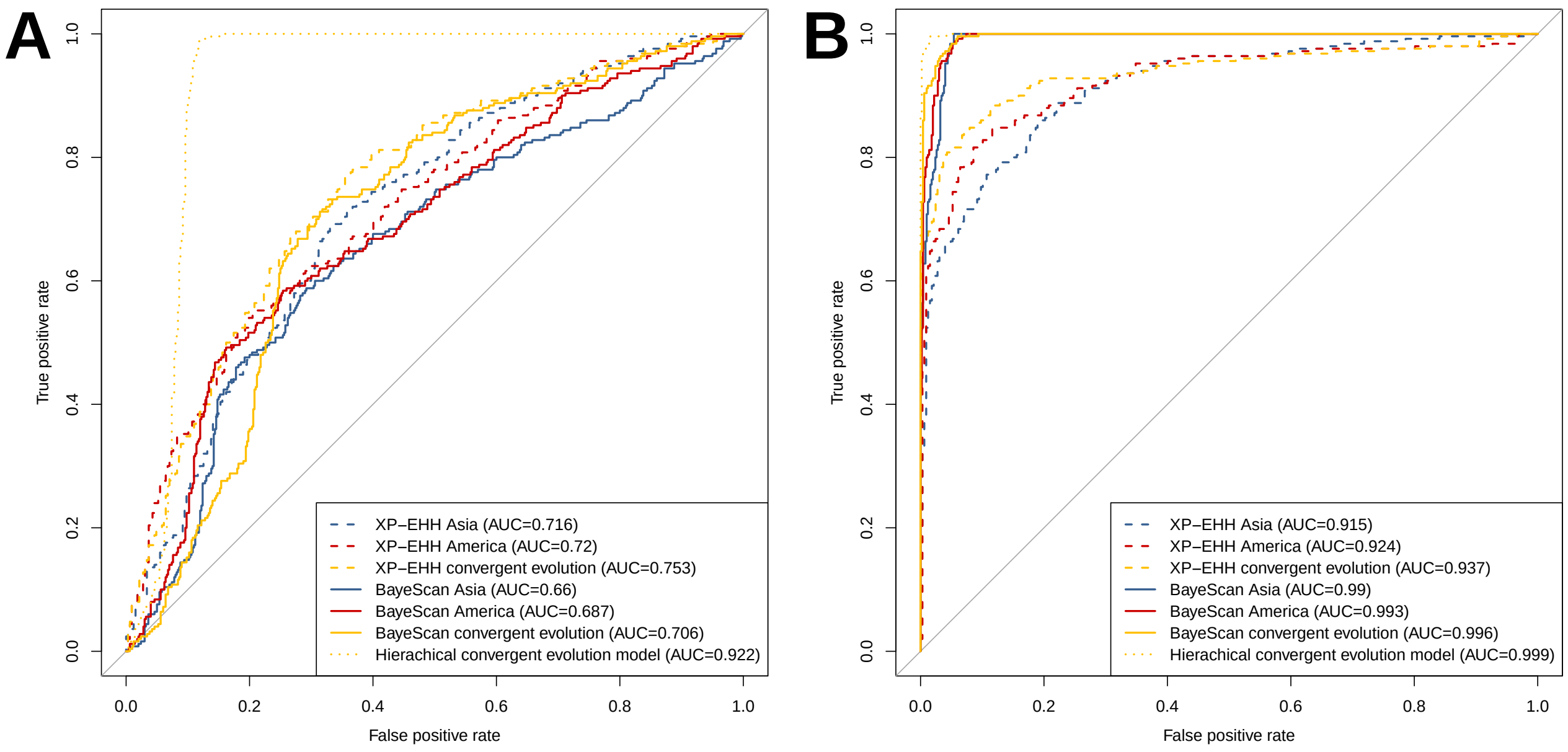
A
True positive rate
False positive rate
0.0
0.2
0.4
0.6
0.8
1.0
XP–EHH Asia (AUC=0.716)
XP–EHH America (AUC=0.72)
XP–EHH convergent evolution (AUC=0.753)
BayeScan Asia (AUC=0.66)
BayeScan America (AUC=0.687)
BayeScan convergent evolution (AUC=0.706)
Hierachical convergent evolution model (AUC=0.922)
B
True positive rate
False positive rate
XP–EHH Asia (AUC=0.915)
XP–EHH America (AUC=0.924)
XP–EHH convergent evolution (AUC=0.937)
BayeScan Asia (AUC=0.99)
BayeScan America (AUC=0.993)
BayeScan convergent evolution (AUC=0.996)
Hierachical convergent evolution model (AUC=0.999)

**Figure S1.**

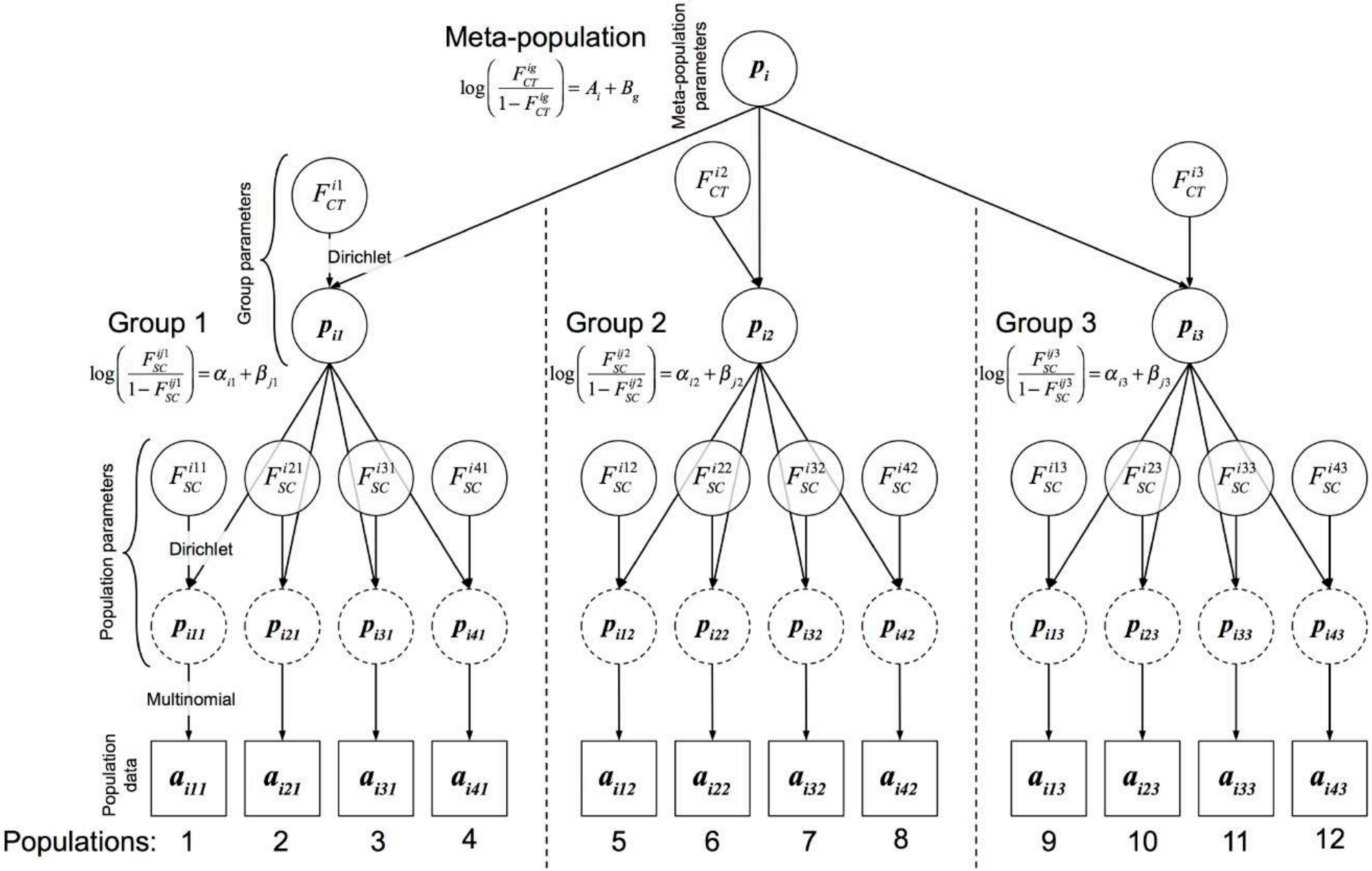

Meta-population
$\log\left(\frac{F_{CT}^{ig}}{1-F_{CT}^{ig}}\right) = A_i + B_g$
Meta-population parameters
$p_i$
Group parameters
$F_{CT}^{i1}$
Dirichlet
$F_{CT}^{i2}$
$F_{CT}^{i3}$
Group 1
$p_{i1}$
$\log\left(\frac{F_{SC}^{ij1}}{1-F_{SC}^{ij1}}\right) = \alpha_{i1} + \beta_{j1}$
Group 2
$p_{i2}$
$\log\left(\frac{F_{SC}^{ij2}}{1-F_{SC}^{ij2}}\right) = \alpha_{i2} + \beta_{j2}$
Group 3
$p_{i3}$
$\log\left(\frac{F_{SC}^{ij3}}{1-F_{SC}^{ij3}}\right) = \alpha_{i3} + \beta_{j3}$
Population parameters
$F_{SC}^{i11}$ $F_{SC}^{i21}$ $F_{SC}^{i31}$ $F_{SC}^{i41}$ $F_{SC}^{i12}$ $F_{SC}^{i22}$ $F_{SC}^{i32}$ $F_{SC}^{i42}$ $F_{SC}^{i13}$ $F_{SC}^{i23}$ $F_{SC}^{i33}$ $F_{SC}^{i43}$
Dirichlet
$p_{i11}$ $p_{i21}$ $p_{i31}$ $p_{i41}$ $p_{i12}$ $p_{i22}$ $p_{i32}$ $p_{i42}$ $p_{i13}$ $p_{i23}$ $p_{i33}$ $p_{i43}$
Multinomial
Population data
$a_{i11}$ $a_{i21}$ $a_{i31}$ $a_{i41}$ $a_{i12}$ $a_{i22}$ $a_{i32}$ $a_{i42}$ $a_{i13}$ $a_{i23}$ $a_{i33}$ $a_{i43}$
Populations: 1 2 3 4 5 6 7 8 9 10 11 12

**Figure S2.**

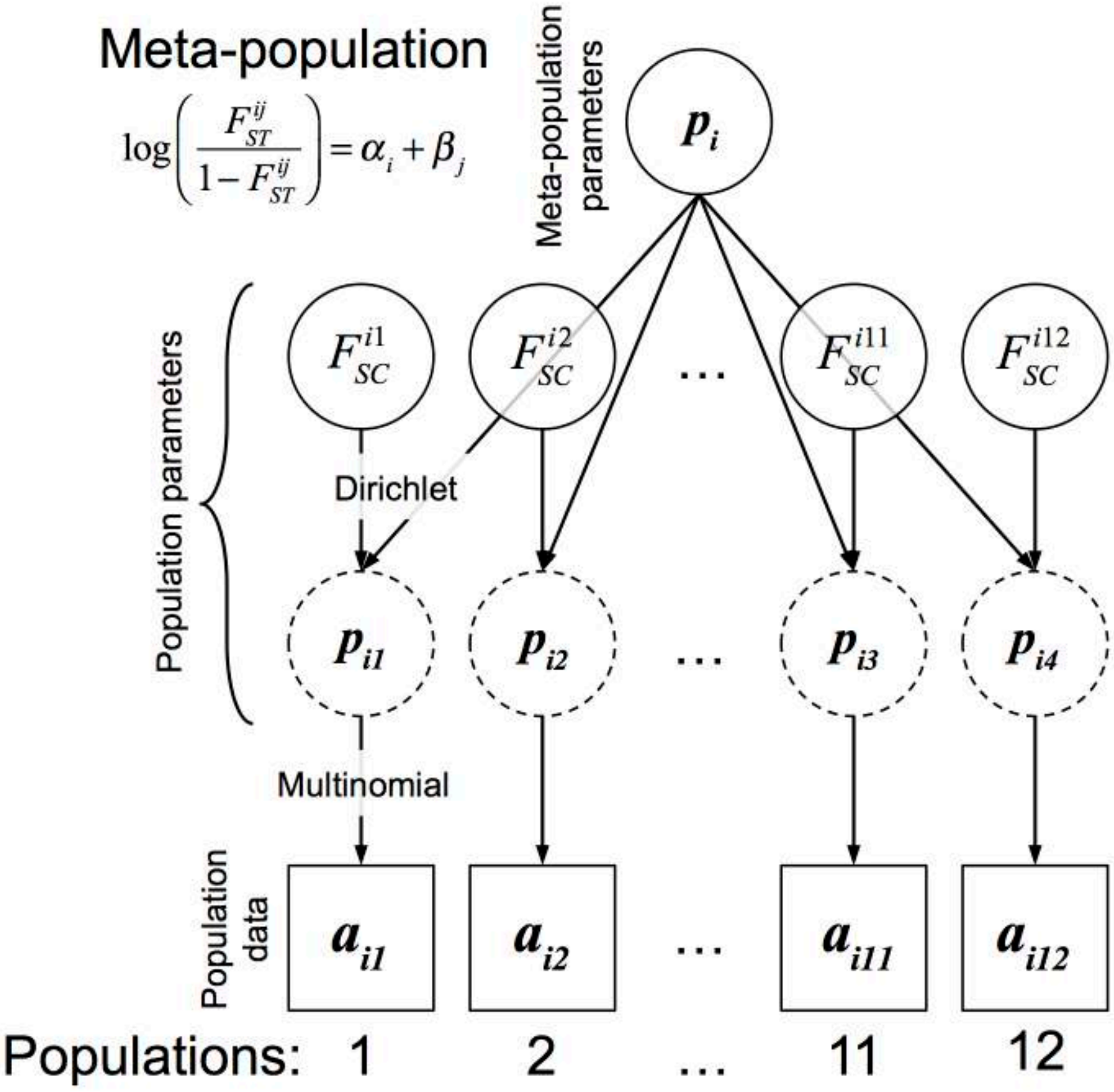
Meta-population
$\log\left(\frac{F_{ST}^{ij}}{1-F_{ST}^{ij}}\right)=\alpha_i+\beta_j$
Meta-population parameters
$p_i$
Population parameters
$F_{SC}^{i1}$
$F_{SC}^{i2}$
...
$F_{SC}^{i11}$
$F_{SC}^{i12}$
Dirichlet
$p_{i1}$
$p_{i2}$
$p_{i3}$
$p_{i4}$
Multinomial
Population data
$a_{i1}$
$a_{i2}$
$a_{i11}$
$a_{i12}$
Populations: 1 2 ... 11 12

## Figure S3.

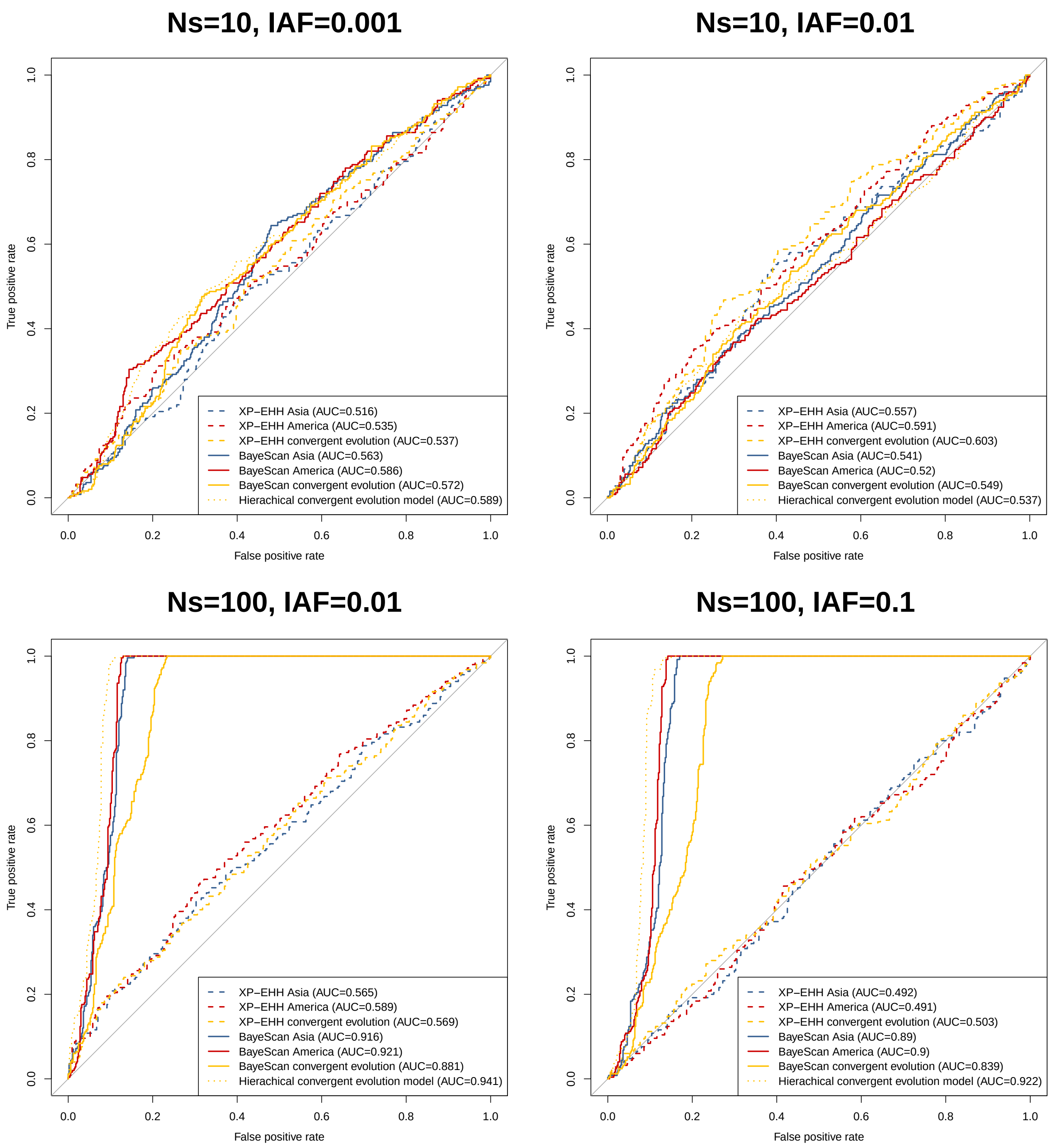
Ns=10, IAF=0.001
Ns=10, IAF=0.01
Ns=100, IAF=0.01
Ns=100, IAF=0.1
True positive rate
False positive rate
0.0
0.2
0.4
0.6
0.8
1.0
XP−EHH Asia (AUC=0.516)
XP−EHH America (AUC=0.535)
XP−EHH convergent evolution (AUC=0.537)
BayeScan Asia (AUC=0.563)
BayeScan America (AUC=0.586)
BayeScan convergent evolution (AUC=0.572)
Hierachical convergent evolution model (AUC=0.589)
XP−EHH Asia (AUC=0.557)
XP−EHH America (AUC=0.591)
XP−EHH convergent evolution (AUC=0.603)
BayeScan Asia (AUC=0.541)
BayeScan America (AUC=0.52)
BayeScan convergent evolution (AUC=0.549)
Hierachical convergent evolution model (AUC=0.537)
XP−EHH Asia (AUC=0.565)
XP−EHH America (AUC=0.589)
XP−EHH convergent evolution (AUC=0.569)
BayeScan Asia (AUC=0.916)
BayeScan America (AUC=0.921)
BayeScan convergent evolution (AUC=0.881)
Hierachical convergent evolution model (AUC=0.941)
XP−EHH Asia (AUC=0.492)
XP−EHH America (AUC=0.491)
XP−EHH convergent evolution (AUC=0.503)
BayeScan Asia (AUC=0.89)
BayeScan America (AUC=0.9)
BayeScan convergent evolution (AUC=0.839)
Hierachical convergent evolution model (AUC=0.922)

## Table S1.

| Chromosome | Region ID | Closest gene to significant SNPs (<250kb) | SNPs in region with qvalue<0.01 |
|---|---|---|---|
| 1 | 1 | ZBTB48 | rs731024 rs6682920 |
| | | KLHL21 | rs6678681 |
| | | THAP3 | rs742394 |
| | | DNAJC11 | rs200457 rs12137794 rs7549198 rs156985 rs3827729 rs399242 rs11122119 rs277671 rs277681 |
| | 2 | - | - |
| | **3** | TRIM67 | rs12096847 rs12063614 rs12563076 rs11122250 rs6541261 |
| | | **TSNAX-DISC1** | rs16854592 rs12058372 rs12058117 |
| 2 | **4** | PRKCE | rs2594489 rs8179696 |
| | | **EPAS1** | rs17034920 rs2121266 rs17034950 rs7571879 rs4952819 rs7582701 rs13419896 rs9679290 rs6758592 rs6715787 rs6544888 rs6756667 rs7589621 rs4953360 rs6755594 rs1374749 rs10178633 rs11675232 rs4953361 rs3088359 rs11678465 rs7583392 rs7594278 rs7571218 rs13006131 rs1868092 rs13424253 rs4450662 rs1109286 rs1447563 rs2346177 rs4953372 rs7587138 |
| | | LOC388946 | rs7556828 rs13003074 rs12622818 rs12619696 rs1542271 rs13032473 rs4953385 rs2276555 rs4953388 rs13001507 rs6741821 |
| | | ATP6V1E2 | rs11125079 rs4953396 rs12470532 rs3814045 |
| | | RHOQ | rs11676473 rs17035292 |
| | | PIGF | rs1901263 rs4953402 rs13000706 |
| | | CRIPT | rs6735530 rs8390 rs3087822 rs12104572 rs12105006 rs7599097 rs10514802 rs10179861 rs10204096 rs4952838 |
| | | SOCS5 | rs12328738 rs6742593 rs991861 rs10209278 rs2346877 rs1319498 rs1378763 rs930853 rs10178013 rs10495936 rs880292 rs7596521 rs7584870 rs7562173 rs11695058 |
| | 5 | SP140L | rs13012615 rs12694851 rs12694855 rs13383946 rs12694858 rs10184953 rs4364013 rs4577270 |
| 3 | 6 | SATB1 | rs336614 rs453585 rs4130090 rs11128893 rs13084808 rs7633180 rs6799759 rs12106877 |
| 4 | **7** | TSPAN5 | rs10031904 rs2178125 rs7685402 |
| | | EIF4E | rs1373244 rs7684429 |
| | | METAP1 | rs17595102 |
| | | **ADH5** | rs1869458 rs2851275 |
| | | **ADH4** | rs3805322 |
| | | **ADH6** | rs2051428 |
| | | **ADH1A** | rs6839510 rs1229966 |
| | | **ADH1B** | rs3811802 |
| | | **ADH7** | rs1442488 rs969804 rs284793 rs284789 rs284787 |
| 5 | 8 | DNAH5 | rs2034221 rs1445691 rs2166337 rs1017573 rs2896110 rs30171 rs2652768 rs1354191 |
| | 9 | CEP120 | rs890933 |
| | | CSNK1G3 | rs11959808 rs7712330 rs7714655 rs10038345 rs12515732 rs11241705 rs10478591 |
| 6 | 10 | ZFP57 | rs3131847 rs3129045 rs2747442 rs3117294 |
| | | HLA-G | rs3115628 |
| | | HLA-A | rs2523969 rs2523957 rs2256919 |
| | | HLA-J | rs2735069 |
| | | RNF39 | rs7382061 |
| | | TRIM31 | rs6457144 rs9261394 rs4959041 |
| | | TRIM40 | rs2517592 rs9261442 rs9261446 rs1541270 rs9261471 rs2857435 rs2857439 rs9261488 rs9261489 rs9261491 rs757262 rs757259 rs1573298 rs9261518 rs9261519 |
| | | TRIM15 | rs9261539 |
| | | TRIM26 | rs1042338 rs3132671 rs2844775 rs3130391 rs3132666 |
| | | HLA-L | rs2844780 |
| 7 | **11** | **IL6** | rs2961299 rs2905324 rs1006001 rs2961304 rs2961309 rs1548418 rs2961312 rs6946864 rs6969502 rs4719711 rs1404008 rs6461662 rs6963591 rs1880242 rs2066992 rs2069852 rs7802277 |
| | | RPS26 | rs4722175 rs9639435 rs9639436 |
| | 12 | ABCB1 | rs1045642 rs6949448 rs4148738 rs10808072 rs2235033 rs1202169 rs1202168 rs1202184 |
| | 13 | OPN1SW | rs1868774 |
| | | IRF5 | rs4728142 rs3807306 |
| | | TNPO3 | rs12531711 rs12531054 rs17424602 |
| | | TPI1P2 | rs12537496 rs13232316 rs17340542 rs13227095 |
| 8 | 14 | PSCA | rs9297976 |
| | | LY6K | rs2164308 rs2082801 rs1469811 rs10956986 |
| | | GML | rs2717586 rs439747 |
| 9 | 15 | TGFBR1 | rs868 |
| | | SEC61B | rs894674 rs920771 rs7032399 rs7040144 |
| 10 | 16 | ARHGAP19 | rs793519 |
| | | RRP12 | rs6584123 |
| | | MMS19 | rs872106 |
| | | UBTD1 | rs10882949 rs10882950 rs10882951 |
| | 17 | NOLC1 | rs7897 |
| | | C10orf26 | rs2250580 rs2249845 rs7069489 rs10786708 rs7079231 rs549466 rs630185 rs2482496 rs2254093 rs2482506 |
| 11 | 18 | RAPSN | rs17198158 |
| | | NDUFS3 | rs4147730 |
| | | MTCH2 | rs4752786 |
| | | AGBL2 | rs4752791 |
| | | NUP160 | rs2305982 rs6485788 rs4752797 rs7924699 rs1872167 |
| 12 | **19** | CUX2 | rs933307 rs1362006 rs7300860 |
| | | SH2B3 | rs739496 |
| | | ATXN2 | rs1029388 |
| | | ACAD10 | rs11066019 |
| | | **ALDH2** | rs4767944 |
| | | MAPKAPK5 | rs4346023 |
| | 20 | CCDC64 | rs11829349 rs11064983 rs7302874 rs12311327 |
| 15 | 21 | OCA2 | rs4778210 rs1800414 rs12593141 rs2305252 rs3794602 rs3829488 rs16950821 rs895828 rs7179419 |
| | | HERC2 | rs916977 |
| 16 | 22 | TOX3 | rs3104767 rs3112625 rs12929797 rs3104780 |
| | | TPM3 | rs3104784 rs3104800 rs3112609 |
| 22 | 23 | TNRC6B | rs8138982 rs12485003 rs17001819 |
| | | MKL1 | rs133054 |
| X | 24 | SMEK3P | rs2218675 rs6418587 rs12859748 rs4240089 rs7061153 rs4297201 rs4357455 rs4829056 |
| | 25 | PLAC1 | rs5933443 rs5933446 rs5930658 rs5978032 rs5930660 |
| | | FAM122B | rs5933454 rs5933455 rs2355307 rs13440516 |

## Table S2.

| Rank | Set size | p-value | q-value | Set name |
|---|---|---|---|---|
| **1** | **15** | **2.00E-06** | **0.001087** | **Fatty Acid Omega Oxidation** |
| 2 | 10 | 2.00E-06 | 0.001087 | Ethanol oxidation |
| 3 | 78 | 2.20E-05 | 0.006520 | Metabolism of xenobiotics by cytochrome P450 |
| 4 | 76 | 2.40E-05 | 0.006520 | Chemical carcinogenesis |
| 5 | 131 | 6.20E-05 | 0.009160 | Biological oxidations |
| **6** | **62** | **7.60E-05** | **0.009160** | **Interferon gamma signaling** |
| 7 | 23 | 9.20E-05 | 0.009160 | Antigen Presentation: Folding, assembly and peptide loading of class I MHC |
| 8 | 49 | 1.00E-04 | 0.009160 | Autoimmune thyroid disease |
| 9 | 37 | 1.00E-04 | 0.009160 | Tyrosine metabolism |
| 10 | 35 | 1.02E-04 | 0.009160 | Graft-versus-host disease |
| 11 | 58 | 1.06E-04 | 0.009160 | Interferon alpha/beta signaling |
| 12 | 34 | 1.14E-04 | 0.009160 | Allograft rejection |
| 13 | 43 | 1.16E-04 | 0.009160 | Intestinal immune network for IgA production |
| 14 | 165 | 1.18E-04 | 0.009160 | Herpes simplex infection |
| 15 | 64 | 1.36E-04 | 0.009853 | Retinol metabolism |
| 16 | 66 | 1.46E-04 | 0.009916 | Phase 1 - Functionalization of compounds |
| 17 | 153 | 2.00E-04 | 0.012785 | Interferon Signaling |
| 18 | 41 | 2.88E-04 | 0.016953 | Fatty acid metabolism |
| 19 | 60 | 3.04E-04 | 0.016953 | Antigen processing and presentation |
| 20 | 945 | 3.12E-04 | 0.016953 | Immune System |
| 21 | 70 | 3.60E-04 | 0.018630 | Drug metabolism - cytochrome P450 |
| 22 | 42 | 3.98E-04 | 0.019660 | Heart Development |
| 23 | 17 | 9.16E-04 | 0.043017 | Estrogen metabolism |
| 24 | 557 | 9.50E-04 | 0.043017 | Adaptive Immune System |
| 25 | 10 | 1.08E-03 | 0.047121 | Abacavir transport and metabolism |

## Table S3.

| Glycolysis / Gluconeogenesis | Ascorbate and aldarate metabolism | Fatty acid metabolism | Tyrosine metabolism | Retinol metabolism | Metabolism of xenobiotics by cytochrome P450 | Drug metabolism - cytochrome P450 | Biological oxidations | Phase 1 - Functionalization of compounds | Ethanol oxidation | Estrogen metabolism | Fatty Acid Omega Oxidation | Chemical carcinogenesis | Gene symbol | Chromosome | SNP | Distance to gene (bp) | 1-qvalue |
|---|---|---|---|---|---|---|---|---|---|---|---|---|---|---|---|---|---|
| | | | | | | | x | | | | | | MAT1A | 10 | rs1143694 | 0 | 0.85055 |
| x | x | x | | | | | x | x | x | | x | | ALDH2 | 12 | rs10744777 | 0 | 0.79765 |
| x | | x | x | x | x | x | x | x | x | | x | x | ADH4 | 4 | rs3805322 | 0 | 0.72808 |
| x | | x | x | x | x | x | x | x | x | | x | x | ADH6 | 4 | rs2051428 | 609 | 0.67424 |
| x | | x | x | x | x | x | x | x | x | | x | x | ADH1B | 4 | rs1229982 | 1360 | 0.6567 |
| x | | x | x | x | x | x | x | x | x | | x | x | ADH1A | 4 | rs4699738 | 21993 | 0.60153 |
| | | | | x | x | x | x | x | | | | x | CYP2A13 | 19 | rs1709086 | 646 | 0.59507 |
| | | | | | x | | x | x | | | | | CYP2F1 | 19 | rs1631814 | 1540 | 0.55743 |
| | | | | | | | x | x | | | | | CYP17A1 | 10 | rs11191416 | 7626 | 0.52756 |
| | | | | | | | x | | | x | | x | SULT1A1 | 16 | rs1968752 | 0 | 0.52476 |
| | | | x | | | | | | | | | | GOT1 | 10 | rs11190103 | 39652 | 0.52083 |
| | | | | x | | | | | | | | | DHRS3 | 1 | rs10779770 | 21214 | 0.50015 |
| | | | | x | x | | x | x | | x | x | x | CYP1A1 | 15 | rs2470891 | 7456 | 0.49364 |
| x | | x | x | x | x | x | x | x | x | | x | x | ADH7 | 4 | rs284787 | 0 | 0.48804 |
| x | | x | x | x | x | x | | | | | | x | ADH5 | 4 | rs17595186 | 3416 | 0.47672 |
| | | | x | | | | | | | | | | METTL6 | 3 | rs1869853 | 0 | 0.47074 |
| | | | x | | | x | x | x | | | | | MAOA | X | rs6323 | 0 | 0.41788 |
| | | x | | | | | | | | | | | ACSL1 | 4 | rs10471180 | 0 | 0.4168 |
| | | | | | x | | | | | | | | AKR1C1 | 10 | rs11252861 | 4236 | 0.40535 |
| | | | | | | | x | | | | | | GCLC | 6 | rs9474542 | 44210 | 0.39784 |
| | | | | | | | x | | | | | | SULT1C2 | 2 | rs4149425 | 0 | 0.39518 |
| x | | | | | | | | | | | | | ENO2 | 12 | rs11064467 | 0 | 0.3932 |
| | | | | | | | | | | x | | | NQO1 | 16 | rs2965753 | 0 | 0.38417 |
| | | | x | | | | | | | | | | GOT2 | 16 | rs4485354 | 33029 | 0.38235 |
| x | | | | | | | | | | | | | PFKP | 10 | rs4881060 | 35423 | 0.37467 |
| | | | | | | | | | | | | x | CHRNA7 | 15 | rs11635209 | 0 | 0.37021 |
| | | | | | x | x | x | | | | | x | GSTA5 | 6 | rs7748890 | 0 | 0.36965 |
| | x | | | x | x | x | x | | | | | x | UGT1A10 | 2 | rs4663965 | 0 | 0.36947 |
| | x | | | x | x | x | x | | | | | x | UGT1A8 | 2 | rs4663965 | 0 | 0.36947 |
| | x | | | x | x | x | x | | | | | x | UGT1A7 | 2 | rs4663965 | 0 | 0.36947 |
| | x | | | x | x | x | x | | | x | | x | UGT1A6 | 2 | rs4663965 | 0 | 0.36947 |
| | x | | | x | x | x | x | | | | | x | UGT1A5 | 2 | rs4663965 | 0 | 0.36947 |
| | x | | | x | x | x | x | | | x | | x | UGT1A9 | 2 | rs4663965 | 0 | 0.36947 |
| | x | | | x | x | x | x | | | | | x | UGT1A4 | 2 | rs4663965 | 0 | 0.36947 |
| | x | | | x | x | x | x | | | x | | x | UGT1A3 | 2 | rs4663965 | 0 | 0.36947 |
| | | | | | | | x | | | x | | | SULT1E1 | 4 | rs1032363 | 34143 | 0.36752 |
| | | | | x | | | x | x | x | | x | | ALDH1A1 | 9 | rs3847322 | 23864 | 0.36343 |
| | | | | | | | x | x | | | | | SMOX | 20 | rs8118315 | 0 | 0.36083 |
| | | | | | x | x | x | | | | | x | GSTA4 | 6 | rs6922246 | 8334 | 0.35653 |
| | | | | | | | x | x | | | | | CYP39A1 | 6 | rs1527687 | 25852 | 0.3554 |
| | | | | | | | | | | | | x | ARNT | 1 | rs16827741 | 0 | 0.33868 |
| x | | | x | | x | x | | | | | | x | ALDH3A1 | 17 | rs4646786 | 0 | 0.33581 |
| x | | | | | | | | | | | | | HK1 | 10 | rs906216 | 0 | 0.3354 |
| | | | | | | | x | x | | | | | CYP46A1 | 14 | rs9324014 | 19671 | 0.31692 |
| | | | | x | | | | | | | | | BCMO1 | 16 | rs7192773 | 0 | 0.30804 |
| x | | | | | | | | | | | | | HKDC1 | 10 | rs10998646 | 3357 | 0.30698 |
| | x | | | x | x | x | | | | | | x | UGT2A3 | 4 | rs2877425 | 38572 | 0.30652 |
| | | | | | | | | | | x | | | ARSD | X | rs9334 | 0 | 0.29825 |
| | | | | | x | | | | | | | x | CBR1 | 21 | rs7278507 | 18169 | 0.29793 |
| | | | | | x | | | | | | | | AKR1C3 | 10 | rs2154305 | 6106 | 0.29609 |
| | | | x | | | | | | | | | | TH | 11 | rs11564708 | 13291 | 0.29563 |
| | x | | | x | x | x | x | | | x | | x | UGT1A1 | 2 | rs4148323 | 0 | 0.28973 |

| | | | | | | | | | | | | | | | | | | |
|---|---|---|---|---|---|---|---|---|---|---|---|---|---|---|---|---|---|---|
| | | | | | | | x | | | | | | SULT2B1 | | 19 | rs2665582 | 0 | 0.28731 |
| x | | | | | | | | | | | | | GALM | | 2 | rs12619390 | 0 | 0.28123 |
| | | | | | | | x | | | | | | GLYAT | | 11 | rs472569 | 14252 | 0.27873 |
| | x | | | x | x | x | x | | | x | | x | UGT2B7 | | 4 | rs12645107 | 10294 | 0.27335 |
| | | x | | | | | | | | | | | ACSL3 | | 2 | rs6756323 | 0 | 0.27255 |
| | | | | | | | x | | | | | | SULT1C4 | | 2 | rs17190797 | 12256 | 0.27007 |
| | | | x | | | | x | | | x | | | COMT | | 22 | rs174699 | 0 | 0.2663 |
| x | | x | x | x | x | x | x | x | x | | x | x | ADH1C | | 4 | rs1614972 | 0 | 0.26505 |
| | | | | | | | x | x | | | | | CYP11B2 | | 8 | rs4736354 | 11509 | 0.26504 |
| | | | | | | | x | x | | | | | CYP24A1 | | 20 | rs17217119 | 27398 | 0.26455 |
| | | | | | x | | | | | | | | CBR3 | | 21 | rs2835288 | 6969 | 0.2632 |
| | | x | | | | | | | | | | | GCDH | | 19 | rs11085825 | 0 | 0.25791 |
| | | | | | x | | | | | | | x | EPHX1 | | 1 | rs17571067 | 0 | 0.25651 |
| | | | | | x | x | x | | | | | x | MGST2 | | 4 | rs8192004 | 221 | 0.25594 |
| x | x | x | | | | | | | | | | | ALDH7A1 | | 5 | rs6862379 | 19248 | 0.25458 |
| x | | | | | | | | | | | | | PGAM1 | | 10 | rs11189214 | 5019 | 0.24525 |
| | | | | | | | x | x | | | | | TBXAS1 | | 7 | rs7806848 | 0 | 0.2445 |
| | | x | | | | | | | | | | | CPT1A | | 11 | rs3019607 | 0 | 0.24162 |
| x | | | | | | | | | | | | | BPGM | | 7 | rs10488453 | 41034 | 0.24073 |
| | | | x | | | | | | | | | | MIF | | 22 | rs4820571 | 5564 | 0.23939 |
| | | | | | | | x | x | | | | | CYP2R1 | | 11 | rs1553006 | 9199 | 0.23262 |
| | x | | | | | | | | | | | | RGN | X | | rs6521178 | 0 | 0.23234 |
| | | x | | | | | | | | | | | ACADM | | 1 | rs1463812 | 8106 | 0.23023 |
| | | x | | | | | | | | | | | ACAT2 | | 6 | rs927450 | 893 | 0.21685 |
| | | | | | | | x | x | | | | | CYP2J2 | | 1 | rs11572223 | 0 | 0.21628 |
| x | | | | | | | | | | | | | PGM2 | | 4 | rs6531588 | 0 | 0.215 |
| | | | x | | | | | | | | | | DDC | | 7 | rs7803788 | 0 | 0.21412 |
| | | | | | x | x | x | | | | | x | GSTP1 | | 11 | rs7952081 | 26133 | 0.2121 |
| | x | | | x | x | x | x | | | | | x | UGT2B11 | | 4 | rs4438816 | 34377 | 0.20968 |
| | | | x | | | | | | | | | | FAHD1 | | 16 | rs3848346 | 0 | 0.209 |
| | | x | | x | | | | | | | | | CYP4A22 | | 1 | rs12564103 | 8172 | 0.20756 |
| x | | | | | | | | | | | | | ADPGK | | 15 | rs11631352 | 10347 | 0.20702 |
| | | | | | | | | | | x | | | STS | X | | rs7884548 | 45611 | 0.20644 |
| | | | | | x | x | x | | | | | x | GSTA2 | | 6 | rs4715309 | 28632 | 0.19774 |
| | | | x | | | x | x | x | | | | | MAOB | X | | rs5952352 | 0 | 0.19713 |
| | | | | x | | | | | | | | | PNPLA4 | X | | rs7888492 | 6078 | 0.19243 |
| | x | | | x | x | x | x | | | | | x | UGT2B10 | | 4 | rs1458232 | 3868 | 0.18923 |
| | | | | | | x | x | x | | | | | FMO3 | | 1 | rs9970392 | 1699 | 0.18754 |
| | | | | | x | | x | x | | x | | x | CYP1B1 | | 2 | rs162556 | 3131 | 0.18684 |
| | | | | | | | x | x | | | | | CYP7B1 | | 8 | rs6472155 | 18859 | 0.18608 |
| x | x | x | | | | | | | | | | | ALDH3A2 | | 17 | rs4646798 | 0 | 0.18088 |
| | | | | | | | x | | | | | | GGT7 | | 20 | rs17092148 | 0 | 0.18042 |
| | | | | | | | x | x | | | | | CYP4F11 | | 19 | rs2072269 | 0 | 0.18022 |
| | | | | | x | | | | | | | x | AKR1C2 | | 10 | rs7915338 | 2307 | 0.17941 |
| | | | | | | | x | x | | | | | CYP4F8 | | 19 | rs3764563 | 1826 | 0.17572 |
| | | | x | | | | | | | | | | IL4I1 | | 19 | rs1290743 | 0 | 0.17491 |
| | | | | x | x | x | x | x | | | | x | CYP2B6 | | 19 | rs16974799 | 0 | 0.17471 |
| | | | | | x | x | x | | | | | x | MGST1 | | 12 | rs10846364 | 49997 | 0.17314 |
| | | | | | | | x | | | | | | GGCT | | 7 | rs38433 | 9205 | 0.17274 |
| | | | x | | | x | | | | | | | AOX1 | | 2 | rs7563911 | 40066 | 0.17263 |
| x | | | | | | | | | | | | | G6PC | | 17 | rs161634 | 0 | 0.17207 |
| | | | | | | | x | | | | | | SULT1B1 | | 4 | rs1847366 | 0 | 0.17034 |
| | | | | | | | x | x | | | | | CYP4F12 | | 19 | rs631193 | 0 | 0.16868 |
| x | | | | | | | | | | | | | ALDOA | | 16 | rs11642740 | 3756 | 0.16786 |
| | | | | | | | x | | | | | | MTR | | 1 | rs4006372 | 10534 | 0.16752 |
| | x | | | x | x | x | x | | | | | x | UGT2B4 | | 4 | rs13151633 | 33826 | 0.166 |
| | x | | | x | x | x | x | | | | | x | UGT2B28 | | 4 | rs13140179 | 10512 | 0.1656 |
| | | | | | | | | | | | | x | CCBL1 | | 9 | rs12551834 | 0 | 0.16554 |
| x | | | | | | | | | | | | | DLD | | 7 | rs6965674 | 3484 | 0.16398 |
| | | | | x | x | x | x | x | | | | x | CYP3A43 | | 7 | rs1403195 | 1305 | 0.16398 |
| | | | x | | | | | | | | | | TPO | | 2 | rs17091737 | 0 | 0.16221 |
| | | | | | | | x | | | | | | MAT2B | | 5 | rs297954 | 40708 | 0.16187 |
| | | | | | | | x | x | | | | | CYP19A1 | | 15 | rs6493496 | 0 | 0.16155 |
| x | | | | | | | x | x | x | | | | ACSS2 | | 20 | rs2295097 | 0 | 0.15989 |
| | | | | | | | x | | | | | | PAPSS1 | | 4 | rs3936010 | 46964 | 0.15733 |
| | | | | | | | x | | | | | | ACSM2B | | 16 | rs11074461 | 9858 | 0.15701 |
| | | x | | | | | | | | | | | ECI2 | | 6 | rs660560 | 0 | 0.15441 |
| x | | | | | | | | | | | | | ENO3 | | 17 | rs238239 | 0 | 0.15396 |
| | | x | | | | | | | | | | | ACOX3 | | 4 | rs1678311 | 29476 | 0.15162 |
| x | x | x | | | | | | | | | | | ALDH1B1 | | 9 | rs7867383 | 1024 | 0.15151 |

| | | | | | | | | | | | | | | | | | |
|---|---|---|---|---|---|---|---|---|---|---|---|---|---|---|---|---|---|
| | | x | | | | | | | | | | ACSL4 | X | | rs5943427 | 0 | 0.15148 |
| | | | | | | | x | | | | | NNMT | | 11 | rs4938091 | 23905 | 0.15101 |
| | | | | | | | | | | | x | CCBL2 | | 1 | rs3753683 | 0 | 0.15083 |
| | | x | | | | | | | | | | ACSBG1 | | 15 | rs1105401 | 0 | 0.14993 |
| x | | | | | | | | | | | | DLAT | | 11 | rs7949136 | 0 | 0.14944 |
| x | | | x | | x | x | | | | | x | ALDH3B1 | | 11 | rs581105 | 0 | 0.14814 |
| | | | | | | | x | | | | | ACSM1 | | 16 | rs433598 | 0 | 0.14765 |
| | | | | | | | x | | | | | SULT4A1 | | 22 | rs138099 | 0 | 0.14752 |
| | | | | | | | x | x | | | | CYP11B1 | | 8 | rs7818826 | 0 | 0.1457 |
| | | | | | x | x | x | | | | x | GSTA3 | | 6 | rs534232 | 7064 | 0.14447 |
| | | | | | | | x | | | | x | NAT1 | | 8 | rs10503610 | 19557 | 0.14389 |
| x | | | | | | | x | x | x | | | ACSS1 | | 20 | rs761720 | 0 | 0.14386 |
| x | | | | | | | | | | | | GAPDH | | 12 | rs7971637 | 256 | 0.14324 |
| | | x | | x | | | x | x | | x | | CYP4A11 | | 1 | rs1502931 | 25714 | 0.14251 |
| | x | | | x | x | x | x | | | | x | UGT2A1 | | 4 | rs6600789 | 43023 | 0.14213 |
| | | | | | | | x | x | | | | POMC | | 2 | rs12619264 | 30244 | 0.1408 |
| | | x | | | | | | | | | | EHHADH | | 3 | rs16859799 | 1700 | 0.14073 |
| | | x | | | | | | | | | | ACSL5 | | 10 | rs1887139 | 13477 | 0.14004 |
| | | | | x | x | x | x | x | | | x | CYP2C8 | | 10 | rs11188183 | 33032 | 0.14004 |
| | | | | x | | | x | x | | | | CYP26B1 | | 2 | rs12477733 | 9326 | 0.13939 |
| | | | | x | | | | | | | | ALDH1A2 | | 15 | rs16977865 | 1211 | 0.13932 |
| | | x | | | | | | | | | | ACAA1 | | 3 | rs2239621 | 0 | 0.13844 |
| | | x | | | | | | | | | | ACAA2 | | 18 | rs7239952 | 29150 | 0.13761 |
| | | | | | x | | | | | | | AKR1C4 | | 10 | rs12247748 | 0 | 0.13696 |
| | | | | | | | x | | | | | PAPSS2 | | 10 | rs3824721 | 0 | 0.13696 |
| x | | | | | | | | | | | | ALDOB | | 9 | rs7467699 | 1985 | 0.13695 |
| | | | | | | | x | | | | | TPMT | | 6 | rs2842934 | 0 | 0.13692 |
| x | | | x | | x | x | | | | | x | ALDH1A3 | | 15 | rs4965339 | 33158 | 0.1368 |
| x | | | | | | | | | | | | G6PC2 | | 2 | rs16856159 | 0 | 0.13671 |
| | | | | | x | | | | | | x | HSD11B1 | | 1 | rs2298930 | 0 | 0.13665 |
| | | | | | | x | x | x | | | | FMO1 | | 1 | rs7551531 | 2783 | 0.13665 |
| | | | x | | | | | | | | | TYR | | 11 | rs10830253 | 0 | 0.13443 |
| | | | | | | | x | x | | | | PTGS1 | | 9 | rs10306150 | 0 | 0.1329 |
| | | x | | | | | | | | | | ACADS | | 12 | rs509152 | 8738 | 0.12971 |
| | | | x | | | | | | | | | HPD | | 12 | rs11043218 | 0 | 0.12971 |
| | | | | x | x | x | x | x | | | x | CYP3A7 | | 7 | rs2687145 | 0 | 0.12842 |
| | | | | x | x | x | x | x | | | x | CYP2C9 | | 10 | rs10509679 | 0 | 0.1278 |
| | | | | x | | | | | | | | SDR16C5 | | 8 | rs4075155 | 17953 | 0.1271 |
| | | x | | | | | | | | | | ACADSB | | 10 | rs10794583 | 0 | 0.12652 |
| | | | | | | x | x | x | | | | FMO2 | | 1 | rs7542361 | 0 | 0.1262 |
| | | | | x | | | | | | | | RDH8 | | 19 | rs4322768 | 23944 | 0.12602 |
| | | | x | | | | | | | | | TAT | | 16 | rs4788811 | 14211 | 0.12535 |
| x | | | | | | | | | | | | PGM1 | | 1 | rs10889433 | 0 | 0.12495 |
| | | | | | x | x | x | | | | x | MGST3 | | 1 | rs16847570 | 26812 | 0.12495 |
| | | | | | | | x | x | | | | CYP4B1 | | 1 | rs12082811 | 38784 | 0.12495 |
| | | | | | x | | x | | | | x | SULT2A1 | | 19 | rs2932766 | 770 | 0.12483 |
| x | | | | | | | | | | | | HK2 | | 2 | rs6732614 | 0 | 0.12437 |
| | | | x | | | | | | | | | TYRP1 | | 9 | rs16929332 | 21693 | 0.1239 |
| | | | | x | | | | | | | | RDH12 | | 14 | rs12882315 | 0 | 0.12343 |
| x | | | x | | x | x | | | | | x | ALDH3B2 | | 11 | rs4930497 | 26916 | 0.1219 |
| | | | | x | | | x | x | | | | CYP26A1 | | 10 | rs7905939 | 3466 | 0.12182 |
| | | | | | x | x | x | | | | x | GSTO2 | | 10 | rs157081 | 0 | 0.12182 |
| | | | | | x | x | x | | | | x | GSTA1 | | 6 | rs9395826 | 11026 | 0.12153 |
| x | | | | | | | | | | | | GCK | | 7 | rs741038 | 0 | 0.12121 |
| | | | | x | | | | | | | | RPE65 | | 1 | rs7515572 | 42939 | 0.12117 |
| | | | | | | | | | | | x | PTGS2 | | 1 | rs16825773 | 19572 | 0.12117 |
| | | | | x | | | | | | | | AWAT2 | X | | rs1409995 | 0 | 0.12044 |
| | x | | | x | x | x | | | | | x | UGT2A2 | | 4 | rs17147521 | 0 | 0.12018 |
| | | | | | | | x | | | | | SLC35D1 | | 1 | rs2815378 | 0 | 0.12005 |
| | | | | | | | x | x | | | | PTGIS | | 20 | rs495146 | 0 | 0.11976 |
| x | | | | | | | | | | | | LDHA | | 11 | rs11024636 | 10445 | 0.11939 |
| | | | | | | | x | | | | | MAT2A | | 2 | rs6705971 | 4684 | 0.11929 |
| | | | | | | | x | x | | | | CYP4F3 | | 19 | rs4807964 | 2880 | 0.11904 |
| | | | | x | | | | | | | | RDH10 | | 8 | rs6472767 | 32248 | 0.11877 |
| x | x | x | | | | | | | | | | ALDH9A1 | | 1 | rs10918241 | 0 | 0.11766 |
| | | | | x | | | | | | | | LRAT | | 4 | rs156500 | 3045 | 0.11752 |
| | | x | | | | | | | | | | ACOX1 | | 17 | rs7217955 | 0 | 0.11751 |
| x | | | | | | | | | | | | PCK1 | | 20 | rs2179706 | 0 | 0.11741 |
| | | | | x | | | | | | | | DHRS4 | | 14 | rs3742492 | 0 | 0.11705 |
| | | | | | x | x | x | | | | x | GSTO1 | | 10 | rs11191966 | 10840 | 0.11699 |

| | | | | | | | | | | | | | | | | |
|---|---|---|---|---|---|---|---|---|---|---|---|---|---|---|---|---|
| | | | x | | | | | | | | | WBSCR22 | 7 | rs10479665 | 0 | 0.11657 |
| x | | | | | | | | | | | | FBP1 | 9 | rs10993280 | 7240 | 0.11643 |
| | | | | x | x | x | x | x | | | x | CYP2C19 | 10 | rs4641393 | 0 | 0.1157 |
| | | | | | x | x | | | | | x | GSTM2 | 1 | rs12024479 | 0 | 0.11429 |
| x | | | | | | | | | | | | FBP2 | 9 | rs497103 | 31419 | 0.11294 |
| | | x | | | | | | | | | | ACSBG2 | 19 | rs4807841 | 0 | 0.11293 |
| | | | | x | x | x | x | x | | | x | CYP2C18 | 10 | rs12777823 | 37749 | 0.1125 |
| | | | | x | | | | | | | | HSD17B6 | 12 | rs11171979 | 14409 | 0.11225 |
| x | | | | | | | | | | | | PGK1 | X | rs2284753 | 0 | 0.11143 |
| x | | | | | | | | | | | | HK3 | 5 | rs691057 | 0 | 0.11125 |
| x | | | | | | | | | | | | LDHB | 12 | rs10841862 | 8920 | 0.11117 |
| | | x | | | | | | | | | | CPT2 | 1 | rs6588471 | 17134 | 0.11088 |
| x | | | | | | | | | | | | PFKL | 21 | rs8134520 | 0 | 0.11027 |
| | | | x | | | | | | | | | DCT | 13 | rs2389137 | 42095 | 0.10951 |
| | | | | | | | x | x | | | | PAOX | 10 | rs11101722 | 2312 | 0.10925 |
| | x | | | | | | x | | | | | UGDH | 4 | rs1450 | 0 | 0.10924 |
| | | x | | | | | | | | | | CPT1B | 22 | rs5770916 | 0 | 0.10914 |
| | | | | | | | x | | | | | BPNT1 | 1 | rs12116864 | 0 | 0.10792 |
| | | x | | | | | | | | | | ACADL | 2 | rs2286963 | 0 | 0.1074 |
| | | | | x | x | x | | | | | x | CYP2A7 | 19 | rs7255901 | 47079 | 0.10739 |
| | | | | | | | x | x | | | | CYP7A1 | 8 | rs16923506 | 12240 | 0.10738 |
| | | | | x | | | | | | | | RDH11 | 14 | rs3759764 | 460 | 0.10722 |
| | | x | | | | | | | | | | HADHB | 2 | rs6745226 | 0 | 0.10635 |
| | | | | x | | | | | | | | RDH16 | 12 | rs7314151 | 0 | 0.10613 |
| | | x | | | | | | | | | | CPT1C | 19 | rs3810265 | 6278 | 0.10551 |
| | | | x | | | | | | | | | FAH | 15 | rs7180031 | 0 | 0.10544 |
| | | | | | | | x | | | | | GGT6 | 17 | rs7216474 | 0 | 0.10544 |
| | | | | x | | | | | | | | DHRS9 | 2 | rs1862069 | 0 | 0.10538 |
| | | | | | x | x | | | | | x | GSTM3 | 1 | rs3814309 | 0 | 0.10523 |
| | | | | x | x | x | x | x | | | x | CYP3A5 | 7 | rs10242455 | 5634 | 0.1049 |
| | | | x | | | | | | | | | GSTZ1 | 14 | rs8177539 | 0 | 0.10464 |
| | | | | | x | | | | | | | AKR7A3 | 1 | rs4483432 | 4683 | 0.10433 |
| x | | | | | | | | | | | | PFKM | 12 | rs11168417 | 0 | 0.10427 |
| | | | | | | | x | x | | | | CYP27A1 | 2 | rs7603709 | 3871 | 0.10358 |
| | | | | | | | x | | | | | UGP2 | 2 | rs4671535 | 0 | 0.10358 |
| | | | x | | | | | | | | | HGD | 3 | rs4676817 | 0 | 0.1035 |
| | | | | | | | x | x | | | | CYP8B1 | 3 | rs6774801 | 6143 | 0.1035 |
| | | x | | | | | | | | | | ACSL6 | 5 | rs82125 | 0 | 0.10296 |
| | | | | | | | x | | | | | GCLM | 1 | rs743110 | 14052 | 0.10273 |
| | | | | | x | x | x | x | | x | | CYP2D6 | 22 | rs5751216 | 13564 | 0.10242 |
| | | | | | | | x | x | | | | CYP4F2 | 19 | rs2074901 | 0 | 0.10214 |
| x | | | | | | | | | | | | PDHA2 | 4 | rs4336232 | 28155 | 0.10155 |
| | | | | | | | x | | | | x | NAT2 | 8 | rs10088180 | 20639 | 0.10124 |
| | | | | | | | x | | | | | CNDP2 | 18 | rs747174 | 8665 | 0.09996 |
| | | | | | | | x | x | | | | CYP51A1 | 7 | rs7797834 | 0 | 0.09992 |
| x | | | | | | | | | | | | PCK2 | 14 | rs973639 | 4677 | 0.09971 |
| | | | x | | | | | | | | | HEMK1 | 3 | rs17787569 | 0 | 0.09964 |
| | | | | x | x | x | x | x | x | x | x | CYP1A2 | 15 | rs2472299 | 7784 | 0.09938 |
| | | | | | | | | | x | | | ARSE | X | rs211641 | 0 | 0.09907 |
| | | | | x | | | | | | | | DGAT1 | 8 | rs3757971 | 0 | 0.09811 |
| x | | | | | | | | | | | | GAPDHS | 19 | rs8100526 | 0 | 0.09762 |
| x | | | | | | | | | | | | PKM | 15 | rs4506844 | 0 | 0.0974 |
| | | | | | x | | | | | | | DHDH | 19 | rs3765148 | 0 | 0.0968 |
| x | | | | | | | | | | | | AKR1A1 | 1 | rs3748645 | 0 | 0.09654 |
| | | | | | | | x | | | | | GGT5 | 22 | rs6004094 | 6698 | 0.09609 |
| x | | | | | | | | | | | | LDHC | 11 | rs35593189 | 0 | 0.0953 |
| | | x | | | | | | | | | | HADHA | 2 | rs2196153 | 0 | 0.09497 |
| | | | x | | | | | | | | | DBH | 9 | rs1611131 | 0 | 0.09399 |
| | | | | | | | x | x | | | | CYP11A1 | 15 | rs3803463 | 7428 | 0.09357 |
| x | | | | | | | | | | | | ENO1 | 1 | rs2765507 | 0 | 0.09329 |
| | | | | | | x | | | | | | FMO4 | 1 | rs1018392 | 42641 | 0.09329 |
| | | | | x | x | x | x | x | | x | x | CYP2A6 | 19 | rs2258314 | 10608 | 0.09317 |
| | | | | | | | x | | | | | GGT1 | 22 | rs140344 | 39562 | 0.09262 |
| | | | | x | | | x | x | | | | CYP26C1 | 10 | rs7073161 | 0 | 0.09248 |
| | | | | | x | x | x | x | | x | x | CYP2E1 | 10 | rs7095379 | 12335 | 0.09248 |
| x | | | | | | | | | | | | PKLR | 1 | rs4620533 | 0 | 0.09236 |
| | | | | x | | | | | | | | RETSAT | 2 | rs999939 | 0 | 0.09188 |
| | | | | | x | | | | | | | AKR7A2 | 1 | rs11808495 | 0 | 0.09141 |
| x | | | | | | | | | | | | PGK2 | 6 | rs12204475 | 6250 | 0.09134 |
| | | | | | | | x | | | | | GSS | 20 | rs6088660 | 0 | 0.09091 |

| | | | | | | | | | | | | | | | | | |
|---|---|---|---|---|---|---|---|---|---|---|---|---|---|---|---|---|---|
| x | | | | | | | | | | | | | GPI | 19 | rs8191425 | 0 | 0.0897 |
| | | | | x | x | x | x | x | | x | x | x | CYP3A4 | 7 | rs6945984 | 6255 | 0.08945 |
| | | | | x | | | | | | | | | RDH5 | 12 | rs3138139 | 0 | 0.08866 |
| | | | | | x | x | | | | | | x | GSTT2 | 22 | rs140245 | 3605 | 0.0885 |
| | | | | | | x | | | | | | | FMO5 | 1 | rs10900326 | 0 | 0.08692 |
| | | | | | x | x | x | | | | | x | GSTM4 | 1 | rs542338 | 5932 | 0.08627 |
| | | | | | | | x | | | | | | AHCY | 20 | rs6088466 | 13926 | 0.08622 |
| | | x | | | | | | | | | | | HADH | 4 | rs221347 | 0 | 0.0856 |
| | | | | | x | x | | | | | | x | GSTT2B | 22 | rs2858908 | 3825 | 0.08455 |
| | | | | | x | x | x | | | | | x | GSTM5 | 1 | rs11807 | 0 | 0.08403 |
| x | | | | | | | | | | | | | LDHAL6A | 11 | rs11024671 | 0 | 0.08301 |
| | | x | | | | | | | | | | | ACAT1 | 11 | rs10890817 | 0 | 0.08278 |
| x | | | | | | | | | | | | | LDHAL6B | 15 | rs3816814 | 0 | 0.08271 |
| | | | | | | | x | x | | | | | CYP27B1 | 12 | rs4646536 | 0 | 0.08271 |
| | | | x | | | | | | | | | | METTL2B | 7 | rs4731470 | 15932 | 0.08121 |
| | | | | | x | | x | x | | | | | CYP2S1 | 19 | rs1645684 | 31842 | 0.08088 |
| | | | | | x | x | | | | | | x | GSTK1 | 7 | rs10248147 | 19499 | 0.08083 |
| | | | | | | | x | x | | | | | CYP2U1 | 4 | rs3756271 | 0 | 0.08076 |
| | | | | x | | | | | | | | | DHRS4L2 | 14 | rs1811890 | 0 | 0.08065 |
| | | | x | | | | | | | | | | AOC2 | 17 | rs16968038 | 46 | 0.08034 |
| | | | | | x | x | x | | | x | | x | GSTM1 | 1 | rs2071487 | 0 | 0.07994 |
| x | | | | | | | | | | | | | PDHB | 3 | rs7231 | 0 | 0.07855 |